\documentclass[12pt, fleqn]{article}
\usepackage{amsmath, amssymb}
\usepackage[dvips]{graphics,color}
\usepackage {graphicx}
\title{Plasma equilibrium equations in coordinates connected with magnetic surfaces. Exact equilibrium solutions.}
\author{Alexei F. Cheviakov\\
\small \emph{Department of Mathematics,
University of British Columbia, V6T 1X7 Canada}\\
\small\textbf{E-mail:} alexch@math.ubc.ca }
\date{}

\newtheorem{statement}{Statement}

\def\const{\hbox{\rm const}}

\def\grad{\mathop{\hbox{\rm grad}}}

\def\max{\mathop{\hbox{\rm max}}}

\def\div{\mathop{\hbox{\rm div}}}
\def\curl{\mathop{\hbox{\rm curl}}}
\def\sn{\hbox{\rm sn}}

\begin{document}

\maketitle
\begin{abstract}
A representation of the static MHD equilibrium system in coordinates connected with magnetic surfaces is suggested. It is
used for producing families of non-trivial 3D exact solutions of isotropic and anisotropic plasma equilibria in different
geometries, with and without dynamics, and often without geometrical symmetries. The ways of finding coordinates in which
exact equilibria can be constructed are discussed; examples and their applications as physical models are presented.
\end{abstract}



\section{Introduction}
\smallskip

In this paper we consider two systems of partial differential equations that are most frequently employed for continuum
description of plasmas.  The isotropic Magnetohydrodynamics (MHD) system has the form {\cite{fusion1}}
\begin{equation}
\frac{\partial\rho}{\partial t}+\div\rho{\bf V}=0\,, \label{eq_MHD_0}
\end{equation}
\begin{equation}
\rho\frac{\partial{\bf{V}}}{\partial t}= \rho{\bf{V}}\times\curl{\bf{V}}- \frac{1}{\mu}\,{\bf{B}}\times\curl{\bf{B}}-\grad P
-\rho\grad\frac{{\bf{V}}^2}2+ \mu_1\triangle{\bf{V}}\,, \label{eq_MHD_1}
\end{equation}
\begin{equation}
\frac{\partial{\bf{B}}}{\partial t}=\curl({\bf{V}}\times{\bf{B}})+ \eta\triangle{\bf{B}}\,,\qquad\eta=
\frac{1}{\sigma\mu}\,, \label{eq_MHD_2}
\end{equation}
\begin{equation}
\div{\bf B}=0\,,\quad{\bf{J}}=\frac{1}{\mu}\curl{\bf{B}}\,. \label{eq_MHD_3}
\end{equation}
Here \textbf{V} is plasma velocity, \textbf{B}, magnetic field, \textbf{J}, electric current density, $\rho$, plasma
density, $P$, pressure, $\mu$, the magnetic permeability of free space, $\sigma$, conductivity coefficient, $\mu_1$, the
plasma viscosity coefficient; $\eta$, resistivity coefficient.

For a vanishing magnetic field, \textbf{B}=0, the above system is reduced to Navier-Stokes equations of motion of a viscous
compressible fluid.

The MHD system correctly describes the medium maintained isotropic by frequent collisions. However, when the mean free path
for particle collisions is long compared to Larmor radius (e.g. in strongly magnetized or rarified plasmas), the
Chew-Golberger-Low (CGL) model {\cite{CGL}} is relevant. Like the MHD system, the CGL equations are derived from Boltzmann
and Maxwell equations, but the density function in Boltzmann equation is expanded in the powers of the Larmor radius. In the
CGL model, the gas pressure $P$ is replaced by a $3\times 3$ pressure tensor
\begin{equation}
{\mathcal{P}}_{ij}=\displaystyle p_\perp \delta_{ij} + \tau B_i B_j, ~~~\tau={\frac{p_\parallel-p_\perp}{\bf{B}^2}}\,,\quad
i,j=1,2,3\,
\end{equation}
with two independent components: the pressure along the magnetic field $p_\parallel$ and in the transverse direction
$p_\perp$.

\smallskip
Under an assumption $\mu_1=\eta=0$, i.e. in the case of non--viscous infinitely--conducting plasmas, both MHD and CGL
systems have several remarkable analytical properties. In particular, one can name, for both systems, the "frozen-in
magnetic field" property (Kelvin's theorem), Lagrangian and Hamiltonian formulation {\cite{newcomb}}, and conservation of
helicity {\cite{woltjer1}}. This approximation is natural in the case of large kinetic and magnetic Reynolds numbers, and is
used in this work.

\smallskip
The present paper is devoted to the study of equilibrium plasma configurations modeled by MHD and CGL equations. The MHD
equilibrium system is
\begin{equation}
\rho{\bf{V}}\times\curl{\bf{V}}- \frac{1}{\mu}\,{\bf{B}}\times\curl{\bf{B}}-\grad P- \rho\grad\frac{{\bf{V}}^2}2=0\,,
\label{eq_MHD_equil1}
\end{equation}
\begin{equation}
\div\bigl(\rho{\bf{V}}\bigr)=0\,,\quad \curl({\bf{V}}\times{\bf{B}})=0\,,\quad \div{\bf{B}}=0\,. \label{eq_MHD_equil2}
\end{equation}

The equilibrium CGL system can be put to the form {\cite{afc_ob, afc_thesis}}
\begin{equation}
\begin{array}{rcl}
\displaystyle\rho{\bf{V}}\times\curl{\bf{V}}- \left(\frac{1}{\mu}-\tau\right){\bf{B}}\times\curl{\bf{B}}&=&
\displaystyle \grad~p_\perp+\rho\grad\frac{{\bf{V}}^2}2+\\
&&\displaystyle+\tau\grad\frac{{\bf{B}}^2}2+ \bf{B}(\bf{B}\cdot\grad\tau)\,, \label{eq_CGL_equil1}
\end{array}
\end{equation}
\begin{equation}
\div\bigl(\rho{\bf{V}}\bigr)=0\,,\quad \curl({\bf{V}}\times{\bf{B}})=0\,,\quad \div{\bf{B}}=0\,. \label{eq_CGL_equil2}
\end{equation}

\smallskip
The above systems are closed by appropriately chosen equations of state (one for MHD equilibria, two for CGL equilibria.) In
this paper we restrict attention to incompressible plasmas
\begin{equation}
\div{\bf{V}}=0. \label{eq_MHD_incompr}
\end{equation}
Incompressibility approximation is commonly used for subsonic plasma flows with low Mach numbers $M\ll1$,
$M^2={\bf{V}}^2/(\gamma P/\rho)$. (Here $\gamma$ is the adiabatic exponent.) For incompressible plasmas, the continuity
equation $\div\rho{\bf{V}}=0$ implies ${\bf{V}}\cdot\grad\rho=0$, hence density is constant on streamlines.

\bigskip
Both of the systems under consideration are non-linear systems of partial differential equations essentially depending on
three spatial variables. No general methods of construction of exact solutions to the corresponding boundary value problems
are available; the question of stability is answered only for particular types of instabilities (for a review, see
\cite{afc_thesis}.) However, some progress have been recently achieved in constructing \emph{particular} exact solutions,
which are also demanded by applications. In this connection, one can mention exact solutions obtained using reductions by
symmetry groups (e.g. Grad-Shafranov and JFKO equations, see {\cite{kadom, obsymm, obsymm3, ball2, ball4, jets4}}) and
solutions constructed using Euler potentials (e.g. {\cite{kaiser2, Mart_Med}}).

\bigskip
The main goal of this paper is to present a more general method of construction of exact solutions to MHD and CGL plasma
equilibrium equations and their static reductions, in different geometries, and with different physical properties.

The method of construction of exact solutions described below is based on the intrinsic property common to both dynamic MHD
and CGL equilibria and many static cases -- the existence of magnetic surfaces {\cite{afc_thesis, obsymm, kk}}.

In many important cases, an \emph{orthogonal} coordinate system can be constructed, with one of the coordinates constant on
magnetic surfaces of some plasma equilibrium configuration. In such coordinates, the static plasma equilibrium system is
reduced to two partial differential equations for two unknown functions. One of the equations of the system is a "truncated"
Laplace equation, and the second has an energy-connected interpretation (Section {\ref{GeomSec}}.)

\smallskip
The suggested representation of the static plasma equilibrium system is used for producing particular exact solutions for
\emph{static and dynamic, isotropic and anisotropic } plasma equilibria in different geometries.

Formulas presented in Section {\ref{AppPropSec}} give rise to explicit expressions for static force-free plasma equilibria
in coordinates with particular relations between metric coefficients. Also, in many classical and non-classical systems of
coordinates, non-trivial \emph{gradient} vector fields can be built, tangent to prescribed sets of magnetic surfaces
(Section {\ref{SolTheorems}}.) Though gradient fields by themselves represent only degenerate plasma equilibria with
constant pressure and no electric currents, they can serve as initial solutions in infinite-parameter transformations (such
as Bogoyavlenskij symmetries {\cite{obsymm, obsymm3}} of Plasma Equilibrium equations, and transformations from MHD to CGL
equilibria {\cite{afc_ob, afc_thesis, afc_denton}}, which produce non-trivial equilibrium configurations in MHD and CGL
framework, with non-vanishing plasma parameters.

\smallskip
In Section {\ref{ExampSec}}, we use the new representation to construct families of MHD and CGL plasma equilibria. We start
from static gradient and non-gradient solutions with magnetic surfaces being nested spheres, ellipsoids, non-circular
cylinders, and surfaces of other types. These solutions, by virtue of Bogoyavlenskij symmetries and MHD-to-CGL equilibrium
transformations, give rise to \emph{infinite families} of more complicated dynamic and static, isotropic and anisotropic
equilibrium configurations. In the majority of constructed equilibria, the behaviour of magnetic energy and other plasma
parameters within the plasma domain is physical.

Auxiliary statements of the Section {\ref{ExtTheorems}} allow, in many cases, the extension of static equilibrium magnetic
fields with a Killing component, thus changing the equilibrium topology. Hence the application of these transformations
before Bogoyavlenskij symmetries results in the change of the domain of arbitrary functions of the latter.

\smallskip
The value of some of the suggested solutions as models of astrophysical phenomena is discussed. It is shown that some
essential features of the models and the relations between macroscopic parameters are in the agreement with astrophysical
observations. Unlike the majority of existing models, the presented solutions are exact and generally non-symmetric.

\section{The representation of Plasma Equilibrium equations in coordinates connected with magnetic surfaces.}\label{GeomSec}

The general MHD equilibrium system (\ref{eq_MHD_equil1})--(\ref{eq_MHD_equil2}), its static reduction
\begin{equation}
\curl{\bf{B}}\times{\bf{B}}=\mu\grad P\,,\quad \div{\bf{B}}=0,\label{eq_PEE}
\end{equation}
and the force-free plasma equilibrium system ($P = \const$)
\begin{equation}
\curl{\bf{B}}=\alpha({\bf{r}}){\bf{B}}\,,\quad \div{\bf{B}}=0\,\label{eq_FF}
\end{equation}
are known to possess a family of 2-dimensional \textit{magnetic surfaces} (or a foliation) $\Psi({\bf r})=\const$, to which
both velocity \textbf{V} and magnetic field \textbf{B} are tangent, and thus magnetic field lines and plasma streamlines lie
on these surfaces\footnote{Magnetic surfaces may not exist only in certain cases of field-aligned dynamic MHD equilibria,
and in the Beltrami case $\curl{\bf B}=\alpha{\bf B}$, $\alpha=\const$. A case-by-case classification is found in
{\cite{afc_thesis}}.} {\cite{obsymm, obsymm3, kk}}.

Noting the value of representation of the static Plasma Equilibrium system (\ref{eq_PEE}) in special coordinates (e.g. the
derivation of Grad-Shafranov and JFKO equations; construction exact of Euler-potential-based solutions), we rewrite this
system of equations in coordinates connected with magnetic surfaces.

Suppose a triply-orthogonal coordinate system $(u,v,w)$ is given, such that the coordinate $w$ enumerates magnetic surfaces
\footnote{The conditions of existence of such systems are discussed in Remark 1 below.}. Then the pressure
$P({\bf{r}})=P(w)$ is constant on magnetic surfaces; the magnetic field has only $u-$ and $v-$ components:
\begin{equation}
{\bf{B}}=B_u{\bf{e}_u} + B_v{\bf{e}_v}. \nonumber
\end{equation}

In orthogonal coordinates the metric tensor is diagonal: $g_{ij}=g_{ii}^2\delta_{ij}$. The usual differential operators in
orthogonal coordinates $(u,v,w)$ have the form
\begin{equation}
\grad f = {\bf{e}}_u \frac{1}{\sqrt{g_{11}}}\frac{\partial f}{\partial u} + {\bf{e}}_v \frac{1}{\sqrt{g_{22}}}\frac{\partial
f}{\partial v} + {\bf{e}}_w \frac{1}{\sqrt{g_{33}}}\frac{\partial f}{\partial w}; \label{eq_GRAD}
\end{equation}
\begin{equation}
\div {\bf{A}} = \frac{1}{\sqrt{g_{11} g_{22} g_{33}}}\left( \frac{\partial }{\partial u}\sqrt{g_{22}} \sqrt{g_{33}} A_1 +
\frac{\partial }{\partial v}\sqrt{g_{11}} \sqrt{g_{33}} A_2 + \frac{\partial }{\partial w}\sqrt{g_{11}} \sqrt{g_{22}}
A_3\right); \label{eq_DIV}
\end{equation}
\begin{equation}
\curl {\bf{A}} = \frac{1}{\sqrt{g_{11} g_{22} g_{33}}}\left| \begin{array}{ccc}
\sqrt{g_{11}} \bf{e_u} & \sqrt{g_{22}} \bf{e_v} & \sqrt{g_{33}} \bf{e_w} \\
\partial/\partial u & \partial/\partial v & \partial/\partial w \\
\sqrt{g_{11}} A_1 & \sqrt{g_{22}} A_2 & \sqrt{g_{33}} A_3 \\
\end{array} \right|
\label{eq_CURL}
\end{equation}
for any differentiable function $f(u,v,w)$ and vector field ${\bf{A}}(u,v,w)$=$A_1 {\bf{e}}_u + $ $A_2 {\bf{e}}_v +$ $A_3
{\bf{e}}_w$.

Using the solenoidality condition $\div{\bf{B}}=0,$ we get

\begin{equation}
\frac{1}{\sqrt{g_{11} g_{22} g_{33}}}\left( \frac{\partial }{\partial u}\sqrt{g_{22}} \sqrt{g_{33}} B_u + \frac{\partial
}{\partial v}\sqrt{g_{11}} \sqrt{g_{33}} B_v\right) =0,\nonumber
\end{equation}
which allows to introduce a potential $F=F(u,v,w)$:
\begin{equation}
\sqrt{g_{22}} \sqrt{g_{33}} B_u = \frac{\partial F}{\partial v},~~~\sqrt{g_{11}} \sqrt{g_{33}} B_v = -\frac{\partial
F}{\partial u}.\nonumber
\end{equation}

Hence the magnetic field is
\begin{equation}
{\bf{B}}=\frac{1}{\sqrt{g_{22} g_{33}}}\frac{\partial F}{\partial v}{\bf{e}_u} - \frac{1}{\sqrt{g_{11}
g_{33}}}\frac{\partial F}{\partial u}{\bf{e}_v}. \nonumber
\end{equation}

Further, the $w$-projection of $\curl{\bf{B}}$ is $0$, as both $\curl{\bf{B}}$ and ${\bf{B}}$ are tangent to magnetic
surfaces $w=\const$. Hence from (\ref{eq_CURL}):
\begin{equation}
\frac{\partial }{\partial u}\left(\frac{\partial F}{\partial u}\frac{\sqrt{g_{22}}}{\sqrt{g_{11} g_{33}}}\right) +
\frac{\partial }{\partial v}\left(\frac{\partial F}{\partial v}\frac{\sqrt{g_{11}}}{\sqrt{g_{22} g_{33}}}\right)=0.
\nonumber
\end{equation}

This is again a potential form: we introduce a potential $\Phi=\Phi(u,v,w)$ by
\begin{equation}
\frac{\partial F}{\partial u}\frac{\sqrt{g_{22}}}{\sqrt{g_{11} g_{33}}} = \frac{\partial \Phi}{\partial v},~~~\frac{\partial
F}{\partial v}\frac{\sqrt{g_{11}}}{\sqrt{g_{22} g_{33}}} = -\frac{\partial \Phi}{\partial u}.\nonumber
\end{equation}

The compatibility conditions for the derivatives of $F$ must be satisfied: $\displaystyle\frac{\partial^2 F}{\partial u
\partial v}=\frac{\partial^2 F}{\partial v \partial u}$. Hence
\begin{equation}
\frac{\partial }{\partial u}\left(\frac{\partial \Phi}{\partial u}\frac{\sqrt{g_{22} g_{33}}}{\sqrt{g_{11}}}\right) +
\frac{\partial }{\partial v}\left(\frac{\partial \Phi}{\partial v}\frac{\sqrt{g_{11} g_{33}}}{\sqrt{g_{22}}}\right) =0.
\label{eq_MHD_nat_sys_1}
\end{equation}

\smallskip
The $u-$ and $v-$components of the equation $\curl{\bf{B}}\times{\bf{B}}=\mu\grad P$ vanish identically; the $w-$component
is
\begin{equation}
\frac{1}{g_{11}}\frac{\partial \Phi}{\partial u}\frac{\partial^2 \Phi}{\partial u \partial w}+\frac{1}{g_{22}}\frac{\partial
\Phi}{\partial v}\frac{\partial^2 \Phi}{\partial v \partial w} = - \mu P'(w).
 \label{eq_MHD_nat_sys_2}
\end{equation}

Hence the system of four static isotropic plasma equilibrium equations (\ref{eq_PEE}), in coordinates $(u,v,w)$ connected
with magnetic surfaces, rewrites as two equations (\ref{eq_MHD_nat_sys_1}), (\ref{eq_MHD_nat_sys_2}).

Using the definition of gradient and laplacian in non-cartesian coordinates (\ref{eq_GRAD}), the new representation can be
written as
\begin{equation}
\triangle_{(u,v)}\Phi=0, \label{eq_MHD_nat_sys_sh1}
\end{equation}
\begin{equation}
{\grad}_{(u,v)}\Phi \cdot {\grad}_{(u,v)} \frac{\partial \Phi}{\partial w} = - \mu P'(w), \label{eq_MHD_nat_sys_sh2}
\end{equation}
where $(u,v)$ means that only $u-$ and $v-$parts of operators are used.

\bigskip
The magnetic field is expressed through the potential $\Phi$ as follows:
\begin{equation}
{\bf{B}}=\frac{1}{\sqrt{g_{11}}}\frac{\partial \Phi}{\partial u}{\bf{e}_u} + \frac{1}{\sqrt{g_{22}}}\frac{\partial
\Phi}{\partial v}{\bf{e}_v} \equiv {\grad}_{(u,v)}\Phi , \label{eq_B_nat_Phi}
\end{equation}
and the electric current:
\begin{equation}
{\bf{J}}=\frac{1}{\mu}\curl{\bf{B}}=\frac{1}{\mu}\left(-\frac{1}{{\sqrt{g_{22} g_{33}}}} \frac{\partial^2 \Phi}{\partial
v\partial w} {\bf{e}_u} + \frac{1}{{\sqrt{g_{11} g_{33}}}} \frac{\partial^2 \Phi}{\partial u\partial w} {\bf{e}_v} \right).
\label{eq_J_nat_Phi}
\end{equation}

\bigskip\noindent \textbf{Remark 1.} A triply orthogonal coordinate system $(u,v,w)$ with $w=\const$ on magnetic surfaces can be constructed not for any
static equilibrium solution $\{{\bf{B}}, P\}$ of (\ref{eq_PEE}). For a family of smooth surfaces $w(x,y,z)=\const$, two
other families $u(x,y,z)=\const$, $v(x,y,z)=\const$ forming a triply orthogonal system can be constructed if and only if the
surfaces $w(x,y,z)=\const$ form \emph{a system of Lam\'{e}}\footnote{It is known (result due to Darboux) that for two
orthogonal families of surfaces to admit the third one orthogonal to both, the two families must intersect in the lines of
curvature. A condition that, for a smooth family $w(x,y,z)=\const$, there exists a second family of surfaces orthogonal to
the given one and intersecting it in the lines of curvature, is found in, e.g., {\cite{eisenhart3}}, and consists of a PDE
of order 3 that $w(x,y,z)$ must satisfy.} {\cite{darboux, eisenhart3}}, i.e., the function $w(x,y,z)$ satisfies a particular
equation of order 3.

There exist many examples of families of Lam\'{e}; they include sets of parallel surfaces; sets of surfaces of revolution;
Ribaucour surfaces, and other families {\cite{eisenhart3}}. Several appropriate examples are discussed below.

\bigskip\noindent \textbf{Remark 2.}
By derivation, every solution $\{\Phi(u,v,w), P(w)\}$ of the system (\ref{eq_MHD_nat_sys_1}) - (\ref{eq_MHD_nat_sys_2}) in
some orthogonal coordinates $(u,v,w)$ defines a static plasma equilibrium with magnetic field (\ref{eq_B_nat_Phi})
satisfying (\ref{eq_PEE}).

\bigskip\noindent \textbf{Remark 3.}
In coordinate systems where $g_{11}=g_{11}(u,v)$, $g_{22}=g_{22}(u,v)$, the second equation of the system,
(\ref{eq_MHD_nat_sys_sh2}), has a simple energy-connected interpretation. Indeed, the equation can be rewritten as
\begin{equation}
\frac{1}{\sqrt{g_{33}}}\frac{\partial}{\partial w}\frac{1}{2}\left({\grad}_{(u,v)}\Phi\right)^2 =
-\frac{\mu}{\sqrt{g_{33}}}P_w, \nonumber
\end{equation}
which is, by (\ref{eq_B_nat_Phi}), equivalent to the relation
\begin{equation}
\frac{1}{\sqrt{g_{33}}}\frac{\partial}{\partial w}\left(\frac{{\bf{B}}^2}{2\mu}+P\right) = 0. \label{eq_energy_grad}
\end{equation}
For incompressible plasma equilibria, the latter means that \emph{the component of the gradient of total energy density in
the direction normal to the magnetic surfaces vanishes}. Therefore for any MHD equilibrium configuration in which magnetic
surfaces $w=\const$ form a family of Lam\'{e}, and where $g_{11}=g_{11}(u,v)$, $g_{22}=g_{22}(u,v)$, \emph{the total energy
can be finite only if the plasma domain is bounded in the direction transverse to magnetic surfaces}.

For example, plasma equilibria found as cylindrically-symmetric solutions of the Grad-Shafranov equation, with domains
unbounded in cylindrical radius $r$ and the polar component of ${\bf{B}}$ vanishing, are available in literature. For such
solutions, in every layer $c_{1}< z < c_{2}$, the total energy is infinite. However, the magnetic energy in layers $c_{1}< z
< c_{2}$ may be finite. The same is true for the solutions obtained in, for example, {\cite{ob_prl, ob_jmp}}.

\bigskip \noindent \textbf{Remark 4.}
As noted by Lundquist {\cite{Lund}}, the static MHD equilibrium equations (\ref{eq_PEE}) are equivalent to the
time-independent incompressible Euler equations that describe ideal fluid equilibria. Therefore static Euler equations may
also be presented in the form (\ref{eq_MHD_nat_sys_1}), (\ref{eq_MHD_nat_sys_2}).

\bigskip \noindent \textbf{Remark 5.}
As will be shown in the sections below, in many cases appropriate orthogonal coordinates $(u,v,w)$ required by the above
theorem may be introduced \emph{globally} in the plasma domain $\mathcal{D}$.

\section{Exact solutions of Plasma Equilibrium equations in the magnetic field-related coordinates and their use for modeling}\label{AppPropSec}

\subsection{Non-trivial Plasma Equilibria arising from in the magnetic field-related coordinate representation}\label{WhatSolsNeeded}

The system of equations (\ref{eq_MHD_nat_sys_1}), (\ref{eq_MHD_nat_sys_2}) under consideration is also essentially
non-linear, and depends on the metric of unknown orthogonal coordinates. In this section, we list several general cases in
which explicit solutions of the representation (\ref{eq_MHD_nat_sys_1}), (\ref{eq_MHD_nat_sys_2}) of Plasma Equilibrium
equations can be found. Appropriate examples are found below.

\bigskip
We are interested in obtaining exact solutions of the system (\ref{eq_MHD_nat_sys_1}), (\ref{eq_MHD_nat_sys_2}) in
different geometries, but do not restrict ourselves to solutions that can be immediately used as models of physical
phenomena. Solutions describing force-free (\ref{eq_FF}) and even "vacuum" magnetic fields ($\curl{\bf{B}}=0,~ P=\const$)
are of interest for physical modeling, because they can serve as \emph{starting solutions} in infinite symmetries and
transformations of Plasma Equilibrium equations.

\smallskip
In particular, it was recently shown {\cite{obsymm, obsymm3}} that ideal incompressible MHD equilibrium equations possess a
Lie group of intrinsic symmetries. If ${\{{\bf{V}}({\bf{r}}),{\bf{B}}({\bf{r}}),P({\bf{r}}),\rho({\bf{r}})\}}$ is a solution
of (\ref{eq_MHD_equil1}), (\ref{eq_MHD_equil2}), (\ref{eq_MHD_incompr}) where the density $\rho({\bf{r}})$ is constant on
both magnetic field lines and streamlines, then
${\{{\bf{V}}_1({\bf{r}}),{\bf{B}}_1({\bf{r}}),P_1({\bf{r}}),\rho_1({\bf{r}})\}}$ is also a solution, where
\begin{equation}
\begin{array}{lll} \displaystyle
{\bf{B}}_1=b({\bf{r}}){\bf{B}}+c({\bf{r}})\sqrt{\mu\rho}\;{\bf{V}}\,,~~~~ \displaystyle
{\bf{V}}_1=\frac{c({\bf{r})}}{a({\bf{r})\sqrt{\mu\rho}}}\,{\bf{B}}+
\frac{b({\bf{r}})}{a({\bf{r}})}\,{\bf{V}}\,,\\
\displaystyle \rho_1=a^2({\bf{r}})\rho,~~~~ P_1=CP+\frac{C{\bf{B}}^2-{\bf{B}}_1
^2}{2\mu}\,,~~~b^2({\bf{r}})-c^2({\bf{r}})=C=\const.
\end{array} \label{eq_OB_symm}
\end{equation}
Here $b({\bf{r}})$, $c({\bf{r}})$ are functions constant on magnetic field lines and plasma streamlines.

\smallskip
A similar Lie group of transformations exists for incompressible CGL equilibria {\cite{afc_thesis, afc_OB}}.

\bigskip
Another class of transformations worth mentioning here is an infinite-parameter map from MHD to CGL equilibrium
solutions {\cite{afc_thesis, afc_denton, afc_OB}}. Let ${\{{\bf{V}}({\bf{r}}), {\bf{B}}({\bf{r}}), P({\bf{r}}),
\rho({\bf{r}})\}}$ be a solution of the system (\ref{eq_MHD_equil1})-(\ref{eq_MHD_equil2}), (\ref{eq_MHD_incompr}) of
incompressible MHD equilibrium equations, where the density $\rho({\bf{r}})$ is constant on both magnetic field lines and
plasma streamlines (i.e. on magnetic surfaces $\Psi=\const$, if they exist.) Then $\{{\bf{V}}_1({\bf{r}}),$
${\bf{B}}_1({\bf{r}}),$ $p_{\perp 1}({\bf{r}}),$
 $p_{\parallel 1}({\bf{r}}),$ $\rho_1({\bf{r}})\}$ is a solution to incompressible CGL plasma equilibrium system
(\ref{eq_CGL_equil1})-(\ref{eq_CGL_equil2}), where
\begin{equation}
{\bf{B}}_1({\bf{r}})  = f({\bf{r}}) {\bf{B}}({\bf{r}}),~~{\bf{V}}_1({\bf{r}})  = g({\bf{r}}) {\bf{V}}({\bf{r}}), ~~\rho _1 =
C_0 \rho({\bf{r}}) \mu / g^2({\bf{r}}),\nonumber
\end{equation}
\begin{equation}
p_{\perp 1}({\bf{r}})=C_0 \mu P({\bf{r}}) + C_1 + (C_0-f^2({\bf{r}})/{\mu}) ~{\bf{B}}^2({\bf{r}}) / 2,\label{eq_MHD_to_CGL}
\end{equation}
\begin{equation}
p_{\parallel 1}({\bf{r}})=C_0 \mu P({\bf{r}}) + C_1 - (C_0-f^2({\bf{r}})/{\mu}) ~{\bf{B}}^2({\bf{r}}) / 2, \nonumber
\end{equation}
and  $f({\bf{r}})$,~ $g({\bf{r}})$ are arbitrary functions constant on the magnetic field lines and streamlines. $C_0, C_1$
are arbitrary constants.

\bigskip
For a given equilibrium solution, in the MHD and CGL frameworks, the above transformations and the Bogoyavlenskij symmetries
can produce a family of solutions connected with it but having different behaviour of physical parameters (pressure, density
and electric current, magnetic and velocity fields). However, such features as solution topology (set of magnetic field
lines and streamlines), stability with respect to certain classes of instabilities, boundedness of energy, are inherited by
the transformed solutions from the original one {\cite{obsymm, obsymm3, afc_thesis, afc_OB}}. If a plasma domain
$\mathcal{D}\in \mathbb{R}^3$ has a boundary, these transformations preserve no-leak-type boundary conditions.

\bigskip
The above-listed transformations can often turn a particular solution into a form suitable for physical modeling. However,
there exists a lack of diversity of exact equilibrium configurations that could serve as "starting points" (this lack is
observed even in the class of static solutions.) Therefore we study the magnetic surface-connected representation of plasma
equilibrium equations (\ref{eq_MHD_nat_sys_1}), (\ref{eq_MHD_nat_sys_2}) aiming at providing methods of building exact
solutions in different geometries, which could later be transformed into families suitable for modeling.

\subsection{Formulas for exact plasma equilibria in particular geometries}\label{SolTheorems}

For a prescribed set of coordinates, the system (\ref{eq_MHD_nat_sys_1}), (\ref{eq_MHD_nat_sys_2}) is a generally nonlinear
system of equations on two unknown functions $\Phi(u,v,w), P(w)$. In coordinates where metric tensor components are
connected in particular ways, formulas defining corresponding plasma equilibria can be explicitly written out.

First we consider cases when $\Phi(u,v,w)$ essentially depends on the magnetic surface coordinate $w$. Thus the plasma
electric current density (\ref{eq_J_nat_Phi}) is nonzero.

\bigskip\textbf{Case (A).}~If the metric tensor components of orthogonal coordinates $(u,v,w)$ satisfy
\begin{equation}
g_{11}/g_{22} = a(u)^2 b(v)^2 c(w)^2 > 0, ~~~g_{33}=\mathcal{F}^2\left(w, \lambda(v) - \mu(u)
\frac{C_2(w)}{c^2(w)C_1(w)}\right), \label{eq_Th_FF_gen_curv1}
\end{equation}
where
\begin{equation}
\displaystyle \frac{d C_1^2(w)}{dw} + c^2(w)\frac{d C_2^2(w)}{dw}=0, \mu(u)=\int{a(u) du}$, $\lambda(v)=\int{1/b(u) du},
\nonumber
\end{equation}
then there exists a solution $\Phi(u,v,w)$ of the system (\ref{eq_MHD_nat_sys_1}), (\ref{eq_MHD_nat_sys_2}) in the form
\begin{equation}
\Phi(u,v,w)=C_1(w)\mu(u)+C_2(w)\lambda(v), \label{eq_Th_FF_gen_curv__phi1}
\end{equation}
with the pressure $P=\const$.

Indeed, under the above assumptions about the relations of $u-$ and $v-$components of the metric tensor, the plasma
equilibrium equations (\ref{eq_MHD_nat_sys_1}), (\ref{eq_MHD_nat_sys_2}) simplify in coordinates $(\mu, \lambda, w)$, and
the solution in the form (\ref{eq_Th_FF_gen_curv__phi1}) is readily found, together with the necessary expression for
$g_{33}$ (\ref{eq_Th_FF_gen_curv1}).

The solution (\ref{eq_Th_FF_gen_curv__phi1}) defines a force-free plasma equilibrium (\ref{eq_FF}) with the proportionality
coefficient
\begin{equation}
\alpha({\bf{r}}) = \alpha(w) = \frac{1}{H_w C_2(w)}\frac{d C_1(w)}{dw}.\nonumber
\end{equation}

\bigskip\textbf{Example.} For spherical coordinates $(r, \theta, \phi)$ = $(w, u, v)$ the metric  $g_{11}=w^2,~~g_{22}=w^2 \sin^2 u,~~g_{33}=1$ satisfies the above
relations (\ref{eq_Th_FF_gen_curv1}). Hence we find a force-free configuration
\begin{equation}
{\bf{B}} = \frac{C_1(r)}{r \sin\theta} {\bf{e}_{\theta}} + \frac{C_2(r)}{r \sin\theta} {\bf{e}_{\phi}},~~\curl{\bf{B}} = \mu
{\bf{J}} =  \alpha(r){\bf{B}},~~ \alpha(r) = \frac{d C_1(r)/dr}{C_2(r)}. \label{eq_simple_sph_sol}
\end{equation}
with spherical magnetic surfaces $r=\const$. Since the magnetic surfaces are rotationally symmetric, several such solutions
can be added to produce a non-symmetric force-free plasma equilibrium tangent to spheres.

Moreover, from the equation (\ref{eq_MHD_nat_sys_2}) it follows that for the spherical case
\begin{equation}
{\bf{B}}^2 = {\frac{1}{w^2}} \left(\left(\frac{\partial\Phi}{\partial u}\right)^2 +
\frac{1}{\sin^2{u}}\left(\frac{\partial\Phi}{\partial v}\right)^2 \right) = - {\frac{1}{w^2}} {\int}_0^w
h^2\frac{dP(h)}{dh}dh + {\frac{1}{w^2}}a_1(u,v), \nonumber
\end{equation}
where $a_1(u,v)$ is generally not identically zero, and is \emph{never} identically zero for force-free plasmas (since
${\bf{B}}^2$ does not vanish in the whole plasma domain). Therefore any force-free and general non-force-free plasma
equilibrium configurations with spherical magnetic surfaces have a pole-type singularity at the origin $r=0$, and infinite
magnetic energy, if the plasma region includes the origin.

\bigskip\textbf{Case (B).}~If the metric tensor components of orthogonal coordinates $(u,v,w)$ satisfy
\begin{equation}
g_{11}/g_{22} = a(u)^2b(v)^2, ~~~g_{33}=\mathcal{F}^2(w), \label{eq_Th_FF_gen_curv2}
\end{equation}
then
\begin{equation}
\Phi_{1}(u,v,w) = \int t(k)\left[C_1(w)e^{\tau n(k)\mu}\cos(n(k)\lambda) + C_2(w)e^{\tau n(k)\mu}\sin(n(k)\lambda)\right]dk,
\label{eq_simple_sol_gen21}
\end{equation}
\begin{equation}
\Phi_{2}(u,v,w) = \int t(k)\left[C_1(w)e^{\tau n(k)\lambda}\cos(n(k)\mu) + C_2(w)e^{\tau n(k)\lambda}\sin(n(k)\mu)\right]dk,
\label{eq_simple_sol_gen22}
\end{equation}
define solutions to the system of isotropic plasma equilibrium system (\ref{eq_MHD_nat_sys_1}), (\ref{eq_MHD_nat_sys_2}),
for any $C_1(w)$ and $C_2(w)$ satisfying $\displaystyle \frac{d}{dw}(C_1^2(w) + C_2^2(w))=0$. Here again $\mu(u)=\int{a(u)
du}$, $\lambda(v)=\int{1/b(u) du}$,~$\tau=\pm 1$,~$n(k)$ is an arbitrary function, and $t(k)$ is an arbitrary generalized
function (for each solution, $n(k), t(k)$ must be chosen so that the integral converges).

These solutions also correspond to force-free plasma equilibria (\ref{eq_FF}) with the coefficient
\begin{equation}
\alpha({\bf{r}}) = \alpha(w) = \frac{1}{\mathcal{F}(w)C_2(w)}\frac{d C_1(w)}{dw}.\nonumber
\end{equation}

\smallskip
This statement follows from noticing that the relations (\ref{eq_Th_FF_gen_curv2}) turn the plasma equilibrium system
(\ref{eq_MHD_nat_sys_1}), (\ref{eq_MHD_nat_sys_2}) into a usual Laplace equation and a total derivative with respect to
$w_1=\int\mathcal{F}(w)dw$:
\begin{equation}
\Phi_{\mu\mu}+\Phi_{\lambda\lambda}=0,~~~\Phi_{\mu}\Phi_{\mu w_1} + \Phi_{\lambda}\Phi_{\lambda w_1} =0; \nonumber
\end{equation}
the solutions (\ref{eq_simple_sol_gen21}), (\ref{eq_simple_sol_gen22}) follow.

\bigskip \noindent \textbf{Remark 1.}
Relations of the type (\ref{eq_Th_FF_gen_curv1}) and (\ref{eq_Th_FF_gen_curv2}) between metric components coefficients are
not unnatural. The simplest example is coordinates obtained by a conformal transformation of the complex plane, as shown in
Section {\ref{subs_eg_gen_cyl}} below.

\bigskip
We now turn attention to the remaining case when the unknown function $\Phi(u,v,w)$ is independent on the magnetic surface
variable $w$.

\bigskip\textbf{Case (C).}~As noted above, the first plasma equilibrium equation (\ref{eq_MHD_nat_sys_1}) in magnetic surface coordinate representation
coincides with the $(u-v)-$part of a Laplacian (\ref{eq_MHD_nat_sys_sh1}). Hence \emph{in any coordinate system where the 3D
Laplace equation $\triangle_{(u,v,w)}\phi(u,v,w)=0$ admits a solution independent of one of the variables ($w$), there
exists a gradient ("vacuum") magnetic field configuration}
\begin{equation}
\div {\bf{B}} = 0,~~\curl {\bf{B}} = 0 \label{eq_lemma_2D_Laplace}
\end{equation}
\emph{corresponding to this solution, and this magnetic field is tangent to surfaces }$w=\const$.

\bigskip \noindent \textbf{Remark 2. Availability of "vacuum" magnetic fields.} Many classical and esoteric coordinate systems
admit geometrically nontrivial two-dimensional solutions of the Laplace equation, as found in literature, for example,
{\cite{moon_sp_fth}}. New systems of coordinates may be constructed where the Laplace's equation will be separable or have
two-dimensional solutions. The list of conditions on the metric coefficients necessary and sufficient for separability of
Laplace equation and existence of two-dimensional solutions is available in \cite{moon_sp_fte}. In the same book one finds
methods of producing new triply orthogonal coordinate systems by conformal transformations of the complex plane.

\bigskip \noindent \textbf{Remark 3. Use of "vacuum" magnetic fields.} Magnetic fields of the type
(\ref{eq_lemma_2D_Laplace}) can be found in different geometries. Under the action of Bogoyavlenskij symmetries
(\ref{eq_OB_symm}) or transformations to anisotropic CGL equilibria (\ref{eq_MHD_to_CGL}), these gradient fields are
transformed into ones with $\curl{\bf{B}}\neq 0$ and give rise to non-trivial dynamic MHD and static and dynamic CGL plasma
equilibria in the same geometry as the original "vacuum" solution. The examples are given below.

\subsection{Construction of additional plasma equilibria using the magnetic surface-related coordinate representation}\label{ExtTheorems}

The following two statements extend classes of solutions of static MHD equilibrium equations
(\ref{eq_MHD_nat_sys_1})-(\ref{eq_MHD_nat_sys_2}) in magnetic-surface-related coordinate representation.  Namely, under
particular conditions on the metric, the equilibrium magnetic field can be extended with a Killing component in the
$w-$direction.

\bigskip
The first statement presents a transformation of a "vacuum" curl-free magnetic vector field depending only on two variables
into an extended magnetic vector field which is \emph{not} force-free or gradient, and thus gives rise to a new
\emph{non-degenerate} solution to plasma equilibrium equations (\ref{eq_PEE}). The example of use of this transformation for
the extention of a solution class is found in the subsequent section {\ref{subs_eg_gen_cyl}}.

\smallskip
\begin{statement}\label{lemma_winding_sol}
If $\phi(u,v)$ is a solution to the system (\ref{eq_MHD_nat_sys_1}),(\ref{eq_MHD_nat_sys_2}) in coordinates $(u,v,w)$ with
properties
\begin{equation}
g_{11}=g_{11}(u,v)=g_{22},~~~g_{33}=g_{33}(w), \label{eq_lemma_winding_sol1}
\end{equation}
then not only the magnetic field (\ref{eq_B_nat_Phi}) with pressure $P(w)=\const$ solves the Plasma Equilibrium equations
(\ref{eq_PEE}), but so does the extended magnetic field
\begin{equation}
{\bf{B}}=\frac{1}{\sqrt{g_{11}}} \frac{\partial \phi}{\partial u} {\bf{e}_{u}} + \frac{1}{\sqrt{g_{22}}} \frac{\partial
\phi}{\partial v} {\bf{e}_{v}} + K(u,v){\bf{e}_{w}} \label{eq_lemma_winding_sol2}
\end{equation}
with pressure
\begin{equation}
P=C-K^2(u,v)/2, \label{eq_lemma_winding_sol_pressure}
\end{equation}
where $K(u,v)$ satisfies
\begin{equation}
\frac{\partial^2 K(u,v)}{\partial u^2}+\frac{\partial^2 K(u,v)}{\partial v^2}=0,~~~\grad\phi(u,v)\cdot\grad
K(u,v)=0.\nonumber
\end{equation}
\end{statement}

This statement is verified directly by substituting the magnetic field (\ref{eq_lemma_winding_sol2}) and the pressure
(\ref{eq_lemma_winding_sol_pressure}) into the static plasma equilibrium system (\ref{eq_PEE}). The function $K(u,v)$ is a
harmonic conjugate of the solution  $\phi(u,v)$.

\bigskip
It turns out to be possible to add a $w-$ component to a wider class of solutions of plasma equilibrium equations
(\ref{eq_MHD_nat_sys_1})-(\ref{eq_MHD_nat_sys_2}) in magnetic-surface-coordinate representation. The following statement
extends a "vacuum" curl-free magnetic vector field depending only on two variables and tangent to surfaces $w=\const$ to
"vacuum" fields that have non-zero $w$-components. The conditions on the metric in this case are more relaxed.

\smallskip
\begin{statement}\label{lemma_ext3}
If $\phi(u,v)$ is a 2-dimensional solution to the plasma equilibrium system
(\ref{eq_MHD_nat_sys_1}),(\ref{eq_MHD_nat_sys_2}) in the coordinates $(u,v,w)$ with properties
\begin{equation}
g_{11}=g_{11}(u,v),~~g_{22}=g_{22}(u,v),~~g_{33}=a(w)^2\mathcal{F}^2(u,v), \label{eq_lemma_ext3}
\end{equation}
then not only the magnetic field (\ref{eq_B_nat_Phi}) with pressure $P(w)=\const$ solves the Plasma Equilibrium equations
(\ref{eq_PEE}), but so does the extended magnetic field
\begin{equation}
{\bf{B}}=\frac{1}{H_u} \frac{\partial \phi}{\partial u} {\bf{e}_{u}} + \frac{1}{H_v} \frac{\partial \phi}{\partial v}
{\bf{e}_{v}} + \frac{D}{H_w}{\bf{e}_{w}};~~~D=\const. \label{eq_lemma_ext3_2}
\end{equation}
This magnetic filed is a vacuum magnetic field: $\div{\bf{B}}=0,~\curl{\bf{B}}=0$ and corresponds to plasma equilibria with
$P=\const$.
\end{statement}

This transformation is used in the example in Section {\ref{sec_spheroids}} below.

\section{Examples of exact plasma equilibria. Physical models.}\label{ExampSec}

\subsection{Isotropic and anisotropic plasma equilibria tangent to nested ellipsoids}\label{sec_halfellips}

In this example, we construct a family of generally non-symmetric plasma equilibria with ellipsoidal magnetic surfaces.

We start from the construction of vacuum magnetic fields tangent to ellipsoids, using the magnetic-surface-connected
representation of plasma equilibria equations (\ref{eq_MHD_nat_sys_1}), (\ref{eq_MHD_nat_sys_2}) (Section
{\ref{SolTheorems}}, Case C.) Then transformations are applied to this gradient solution to produce families of non-trivial
isotropic and anisotropic plasma equilibria.

An application of the resulting solutions to modeling solar photosphere plasma near active regions is discussed.

\bigskip \noindent \textbf{(i). A vacuum magnetic field tangent to ellipsoids.}
The ellipsoidal coordinates are {\cite{moon_sp_fth}}:
\begin{equation}
\begin{array}{ll}
u=\theta, & b^2 < \theta^2 < c^2,\\
v=\lambda, & 0 \leq \lambda^2 < b^2,\\
w=\eta, & c^2 < \eta^2 < +\infty.
\end{array}
\nonumber
\end{equation}

The coordinate surfaces are
\begin{equation}
\begin{array}{lll}
\displaystyle \frac{x^2}{\eta^2}+\frac{y^2}{\eta^2-b^2}+\frac{z^2}{\eta^2-c^2}=1, & ~~($ellipsoids$, \eta=\const),\\
\displaystyle\frac{x^2}{\theta^2}+\frac{y^2}{\theta^2-b^2}-\frac{z^2}{c^2-\theta^2}=1, & ~~($one-sheet hyperboloids$, \theta=\const),\\
\displaystyle\frac{x^2}{\lambda^2}-\frac{y^2}{b^2-\lambda^2}-\frac{z^2}{c^2-\lambda^2}=1, & ~~($two-sheet hyperboloids$,
\lambda=\const).
\end{array}
\nonumber
\end{equation}

Laplace's equation is separable in ellipsoidal coordinates, and we take a solution depending only on $(\theta, \lambda)$, so
that its gradient has zero $\eta$-projection transverse to ellipsoids, but is tangent to them:
\begin{equation}
\Phi_1(\theta, \lambda) = \left(A_1+B_1
{\sn}^{-1}\left(\sqrt{\frac{c^2-\theta^2}{c^2-b^2}},\sqrt{\frac{c^2-b^2}{c^2}}\right)\right)\left(A_2+B_2
{\sn}^{-1}\left(\frac{\lambda}{b},\frac{b}{c}\right)\right). \nonumber
\end{equation}
Here $\sn(x,k)$ is the Jacobi elliptic sine function. The inverse of it is an incomplete elliptic integral
\begin{equation}
F_{ell}(z,k) = \int_0^z{\frac{1}{\sqrt{1-t^2} \sqrt{1-k^2t^2}}~dt}. \nonumber
\end{equation}

$\Phi_1(\theta, \lambda)$ does not depend on $w$, and therefore evidently satisfies both equations (\ref{eq_MHD_nat_sys_1}),
(\ref{eq_MHD_nat_sys_2}). The resulting magnetic field (\ref{eq_B_nat_Phi}) is tangent to ellipsoids $\eta=\const$, and has
a singularity at $\theta=\lambda$, i.e. on the plane $y=0$.

However one may verify that for a plasma region $c<\eta_1<\eta<\eta_2$ the total magnetic energy $\int_V{B^2/2 dv}$ is
finite. Also, if one restricts to a half-space $y>0$ or $y<0$, then the magnetic field is well-defined in a continuous and
differentiable way.

If the magnetic field is tangent to the boundary of a domain, one can safely assume that outside of it ${\bf{B}}=0$
identically. This is achieved, as usual, by introducing a boundary surface current
\begin{equation}
{\bf{i}}_b({\bf{r}}_1)=\mu^{-1}{\bf{B}}({\bf{r}}_1)\times{\bf{n}}_{out}({\bf{r}}_1), \label{eq_bdry_current}
\end{equation}
where ${\bf{r}}_1$ is a point on the boundary of the domain, and ${\bf{n}}_{out}$ is an outward normal.

\begin{figure}[tp]

\begin{center}

\vspace{1 cm}

\scalebox{0.6}{\includegraphics[0cm,0cm][25cm,20cm]{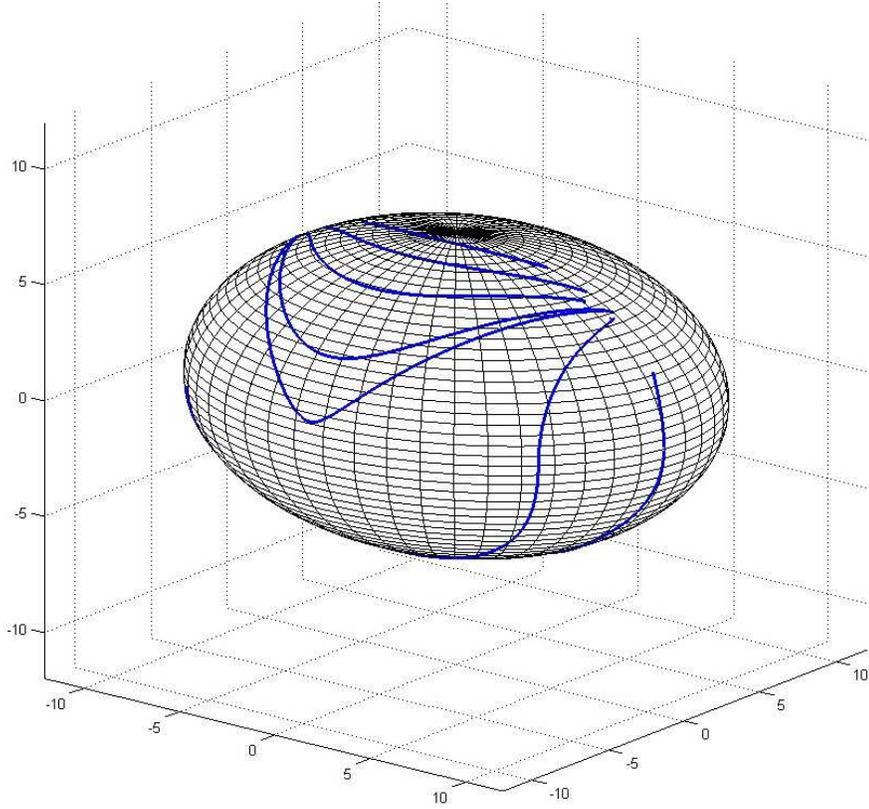}}  

\vspace{3cm}

\end{center}
\renewcommand{\baselinestretch}{1}\small\normalsize

\caption{\label{fig403}A magnetic field tangent to ellipsoids.}{\bigskip Lines of the magnetic field
(\ref{eq_Magn_field_on_ell}) tangent to the ellipsoid $\eta=12$. The shown ellipsoid is a magnetic surface from the family
of nested ellipsoids $\eta=\const$ in classical ellipsoidal coordinates.

This configuration is smoothly defined in the half-space $y>0$.}

\renewcommand{\baselinestretch}{1.8}\small\normalsize
\pagebreak
\end{figure}

\bigskip
Fig. {\ref{fig403}} shows several magnetic field lines for the case $(b=7, c=10, A_1=A_2=0, B_1=1/100, B_2=1/30)$ on the
ellipsoid $\eta=12$. For this set of constants, the vector of the magnetic field has the form is
\begin{equation}
{\bf{B}}_0= \frac{F_{ell}\left(\frac{\lambda}{7},\frac{7}{10}\right)}{\sqrt{(\theta^2-\lambda^2)(\eta^2-\theta^2)}}
{\bf{e}_{\theta}}
 -\frac{F_{ell}\left(\sqrt{\frac{100-\theta^2}{51}},\sqrt{\frac{51}{100}}\right)}{\sqrt{(\eta^2-\lambda^2)(\theta^2-\lambda^2)}}
{\bf{e}_{\lambda}} \label{eq_Magn_field_on_ell}
\end{equation}

This "vacuum" (gradient) magnetic field is used to produce non-trivial dynamic isotropic and anisotropic plasma equilibria,
as shown below.

We remark that though the magnetic field lines of the field (\ref{eq_Magn_field_on_ell}) have a plane of symmetry $x=0$, a
non-zero choice of constants $(A_1, A_2, B_1, B_2)$ would produce a \emph{completely non-symmetric} magnetic field tangent
to a family of ellipsoids.

\bigskip \noindent \textbf{(ii). Isotropic dynamic plasma equilibrium with ellipsoidal magnetic surfaces.}
The above vacuum magnetic field ${\bf{B_0}}$ is indeed a trivial solution to the general isotropic plasma equilibrium system
(\ref{eq_MHD_equil1})-(\ref{eq_MHD_equil2}) with ${\bf{V}}=0,~P=P_0=\const$ and an arbitrary density function
$\rho=\rho_0({\bf{r}})$.

If we choose $\rho_0({\bf{r}})$ to be constant on magnetic field lines (plasma streamlines do not exist as there is no
flow), then the infinite-parameter transformations (\ref{eq_OB_symm}) become applicable to such configuration. Applying them
formally, we obtain a family of isotropic plasma equilibria
\begin{equation}
{\bf{B}}_1  = m({\bf{r}}) {\bf{B}}_0,~~{\bf{V}}_1  = \frac{{n({\bf{r}})}}{{a({\bf{r}})\sqrt {\mu \rho_0({\bf{r}}) } }}
{\bf{B}}_0,\nonumber
\end{equation}
\begin{equation}
\rho _1  = a^2 ({\bf{r}})\rho_0({\bf{r}}),~~P_1  = CP_0 - n^2 ({\bf{r}}) {\bf{B}}_0^2/{(2\mu )}, \label{eq_eg_ell_isotr}
\end{equation}
\begin{equation}
m^2 ({\bf{r}}) - n^2 ({\bf{r}}) = C = \const,\nonumber
\end{equation}
where $a({\bf{r}}), m({\bf{r}}), n({\bf{r}}), \rho_0({\bf{r}})$ are functions constant on magnetic field lines and
streamlines (which coincide in this case, as ${\bf{V}}_1$ and ${\bf{B}}_1$ are collinear).

\bigskip
We consider a plasma configuration in a region $\mathcal{D}$ in the half-space $y>0$ between two ellipsoid shells $\eta_1,~
\eta_2: c<\eta_1<\eta<\eta_2$, (we take $c=10$; $c$ is one of the parameters of the elliptic coordinate systems used for
this solution.)

Outside of the region, we assume ${\bf{B}}_1=0$, by introducing a corresponding surface current ({\ref{eq_bdry_current}}).
We also assume ${\bf{V}}_1=0$, which can be done because the streamlines are tangent to the boundary of the plasma domain
$\mathcal{D}$.

\bigskip
The magnetic field lines in the chosen region are not dense on any 2D surface or in any 3D domain, therefore the arbitrary
functions $a({\bf{r}}), m({\bf{r}}), n({\bf{r}}), \rho_0({\bf{r}})$ can be chosen (in a smooth way) to have a constant value
on each magnetic field line, thus being in fact functions of \emph{two variables} enumerating all magnetic field lines in
the region of interest (for example, $\eta$ and $\lambda$, which specify the beginning of every magnetic field line).

\bigskip
We remark that unlike the initial field ${\bf{B}}_0$, the vector fields ${\bf{B}}_1$ and ${\bf{V}}_1$ are neither potential
nor force-free: for example, $\curl {\bf{B}}_1 = \grad m({\bf{r}})\times{\bf{B}}_0 \nparallel {\bf{B}}_1$. But both
${\bf{B}}_1$ and ${\bf{V}}_1$ satisfy the solenoidality requirement.

\bigskip
Direct verification shows that, with a non-singular choice of the arbitrary functions, the total magnetic energy
$E_m=1/2\int_V{B_1^2~dv}$  and the kinetic energy $E_k=1/2\int_V{\rho~V_1^2~dv}$ are finite. The magnetic field, velocity,
pressure and density $({\bf{B}}_1,~{\bf{V}}_1,~\rho _1,~P_1)$ are defined in a continuous and differentiable way.

\bigskip
The presented model is not unstable according to the known sufficient instability condition for incompressible plasma
equilibria with flows proven in {\cite{fried_vish}} (see also {\cite{afc_thesis}}.) The latter states that if
${\bf{V}}\nparallel {\bf{B}}$, then a plasma equilibrium with constat density is unstable. In the presented example,
${\bf{V}}\parallel {\bf{B}}$ (the density $\rho_1$ can be chosen constant).

\bigskip \noindent \textbf{(iii). Anisotropic plasma equilibrium with ellipsoidal magnetic surfaces.}
When the mean free path for particle collisions is long compared to Larmor radius, (e.g. in strongly magnetized plasmas),
the tensor-pressure CGL approximation should be used. The model suggested here describes a rarefied plasma behaviour in a
strong magnetic field looping out of the star surface.

To construct an anisotropic CGL extension of the above isotropic model, we use the transformations (\ref{eq_MHD_to_CGL})
(Chapter 3) from MHD to CGL equilibrium configurations. Given ${\bf{B}}_1, {\bf{V}}_1, P_1, \rho_1$ determined by
(\ref{eq_eg_ell_isotr}) with some choice of the arbitrary functions $a({\bf{r}})$, $m({\bf{r}})$, $n({\bf{r}})$,
$\rho_0({\bf{r}})$, we obtain an anisotropic equilibrium ${\bf{B}}_2, {\bf{V}}_2, p_{\parallel 2}, p_{\perp 2}, \rho_2$
defined as
\begin{equation}
{\bf{B}}_2  = f({\bf{r}}) {\bf{B}}_1,~~{\bf{V}}_2  = g({\bf{r}}) {\bf{V}}_1, ~~\rho_2 = C_0 \rho_1 \mu /
g({\bf{r}})^2,\nonumber
\end{equation}
\begin{equation}
p_{\perp 2}=C_0 \mu P_1 + C_1 + (C_0-f({\bf{r}})^2/{\mu}) ~{\bf{B}}_1^2 / 2,\label{eq_eg_ell_anisotr}
\end{equation}
\begin{equation}
p_{\parallel 2}=C_0 \mu P_1 + C_1 - (C_0-f({\bf{r}})^2/{\mu}) ~{\bf{B}}_1^2 / 2, \nonumber
\end{equation}
$f({\bf{r}})$,~ $g({\bf{r}})$ are arbitrary functions constant on the magnetic field lines and streamlines, i.e. again on
constant on every plasma magnetic field line, and $C_0, C_1$ are arbitrary constants.

Setting $P_0=0$ in (\ref{eq_eg_ell_isotr}) and making an explicit substitution, we get
\begin{equation}
{\bf{B}}_2  = f({\bf{r}}) m({\bf{r}}) {\bf{B}}_0,~~{\bf{V}}_2  = g({\bf{r}}) \frac{{n({\bf{r}})}}{{a({\bf{r}})\sqrt {\mu
\rho_0({\bf{r}}) } }} {\bf{B}}_0, ~~\rho_2 = C_0 a^2 ({\bf{r}})\rho_0({\bf{r}}) \mu / g({\bf{r}})^2,\nonumber
\end{equation}
\begin{equation}
p_{\perp 2}=C_1 + \frac{{\bf{B}}_0^2}{2\mu}\left(C_0 C \mu - f^2({\bf{r}}) m^2({\bf{r}})\right), \label{eq_eg_ell_anisotr2}
\end{equation}
\begin{equation}
p_{\parallel 2}=C_1 + \frac{{\bf{B}}_0^2}{2\mu}\left( f^2({\bf{r}}) m^2({\bf{r}}) - C_0 C \mu - 2 C_0 n^2({\bf{r}})\right).
\nonumber
\end{equation}

It is known ({\cite{afc_denton, afc_OB}}) that for the new equilibrium to be free from a fire-hose instability, the
transformations (\ref{eq_MHD_to_CGL}) must have $C_0>0$.

$p_{\perp}$ is the pressure component perpendicular to magnetic field lines. It is due to the rotation of particles in the
magnetic field. Therefore in strongly magnetized or rarified plasmas, where the CGL equilibrium model is applicable, the
behaviour of $p_{\perp}$ should reflect that of ${\bf{B}}^2$.

\bigskip
In the studies of the solar wind flow in the Earth magnetosheath, the relation
\begin{equation}
p_\perp/p_\parallel=1+0.847(B^2/(2 p_\parallel)) \label{eq_empiric_anis1}
\end{equation}
was proposed {\cite{empiric1}}. We denote $k({\bf{r}})=C_0 C \mu - f^2({\bf{r}}) m^2({\bf{r}})$ and select the constants and
functions $C_0, C, f({\bf{r}}), m({\bf{r}})$ so that $k({\bf{r}})\geq 0$ in the space region under consideration. From
(\ref{eq_eg_ell_anisotr2}), we have:
\begin{equation}
p_{\perp 2}-p_{\parallel 2}= \frac{{\bf{B}}_0^2}{2\mu} (2 k({\bf{r}}) + 2 C_0 n^2({\bf{r}}) ), \nonumber
\end{equation}
or
\begin{equation}
\frac{p_{\perp 2}}{p_{\parallel 2}}= 1 + \frac{2 k({\bf{r}}) + 2 C_0 n^2({\bf{r}}) } {\mu f^2({\bf{r}}) m^2({\bf{r}})}
 \frac{{\bf{B}}_2^2}{2 p_{\parallel 2}}, \nonumber
\end{equation}
which generalizes and includes the experimental result (\ref{eq_empiric_anis1}).

\bigskip \noindent \textbf{(iv). A model of plasma behaviour in arcade solar flares.}

Solar flares are known to take place in the photospheric region of the solar atmosphere and are connected with a sudden
release of huge energies (typically $10^{22}-10^{25}$ J) (e.g. {\cite{bisk}}, pp. 331-348). Particle velocities connected
with this phenomenon (about $10^3$ m/s) are rather small compared to typical coronal velocities ($\sim 5 \cdot 10^5$ m/s),
therefore equilibrium models are applicable.

Morphologically two types of solar flares are distinguished: loop arcades (magnetic flux tubes) and two-ribbon flares.
Flares themselves and post-flare loops are grounded in from active photospheric regions.

As noted in {\cite{bisk}}, p. 332, \emph{"rigorous theoretical modelling has mainly been restricted to symmetric
configurations, cylindrical models of coronal loops and two-dimensional arcades."}

The configurations described in (ii) and (iii) can serve as \emph{non-symmetric} 3D isotropic and anisotropic models of
quasi-equilibrium plasma in flare and post-flare loops, where magnetic field and inertia terms prevail upon the gravitation
potential term in the plasma equilibrium equations:
\begin{equation}
{\bf{V}}\times{\rm{curl~}}{\bf{V}} \gg \grad \varphi, ~~~ \frac{1}{\mu } {\bf{B}}\times{\rm{curl}}~{\bf{B}} \gg \rho \grad
\varphi.\nonumber
\end{equation}
where $\varphi$ is the star gravitation field potential.

\begin{figure}[tp]

\begin{center}

\vspace{1cm}

\scalebox{0.25}{\includegraphics[0cm,0cm][25cm,20cm]{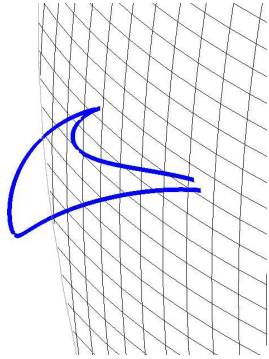}}  

\vspace{3cm}

\end{center}
\renewcommand{\baselinestretch}{1}\small\normalsize

\caption{\label{fig404}A solar flare model -- magnetic field lines.}{\bigskip The model of a solar flare as a coronal plasma
loop near an active photospheric region. The position and shape of several magnetic field lines are shown with respect to
the star surface; magnetic field is tangent to nested half-ellipsoids.}

\renewcommand{\baselinestretch}{1.8}\small\normalsize
\pagebreak
\end{figure}

\begin{figure}[tp]

\begin{center}

\vspace{2cm}

\scalebox{0.4}{\includegraphics[0cm,0cm][25cm,20cm]{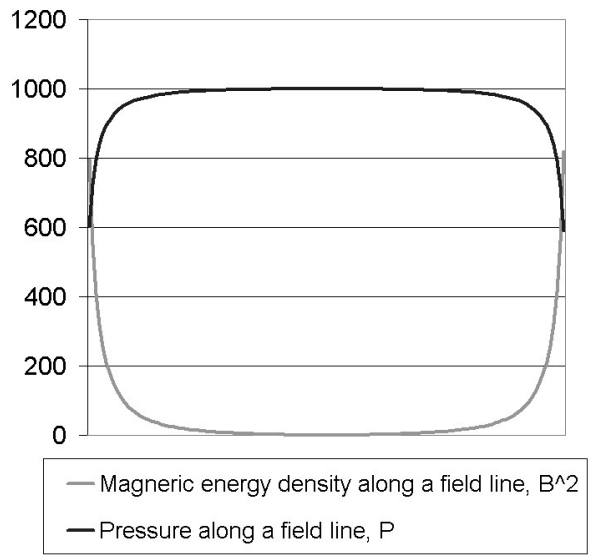}}  

\vspace{3cm}

\end{center}
\renewcommand{\baselinestretch}{1}\small\normalsize

\caption{\label{fig405}A solar flare model -- plasma parameter profiles.}{\bigskip A model of a solar flare -- a coronal
plasma loop near an active photospheric region.

The figure shows the characteristic shape of the magnetic field energy density ${\bf{B}}$ and the pressure $P$ curves along
a particular magnetic field line. (The isotropic case.) }

\renewcommand{\baselinestretch}{1.8}\small\normalsize
\pagebreak
\end{figure}

\bigskip
The relative position and form of the magnetic field lines in the model, with respect to the star surface, are shown on Fig.
{\ref{fig404}}. The characteristic shape of the magnetic field energy density ${\bf{B}}$ and the pressure $P$ along a
particular magnetic field line, for the isotropic case (ii), are given on Fig. {\ref{fig405}}.

\bigskip
Magnetic field lines in the model are not closed; therefore by introducing a surface current of the type
(\ref{eq_bdry_current}), a plasma domain $\mathcal{D}$ can indeed be restricted to any flux tube, with boundary tangent to
magnetic field lines, and the magnetic field can be chosen zero outside (together with the velocity in models (ii), (iii))
by the introduction of a boundary surface current. The current sheet introduction is not artificial -- as argued in
{\cite{parker83}}, in a general 3D coronal configurations the current sheets between flux tubes are formed (see also:
{\cite{bisk}}, p. 343.)

\bigskip
The isotropic MHD model (ii) is valid when the mean free path of plasma particles is much less than the typical scale of the
problem, so that the picture is maintained nearly isotropic via frequent collisions.

However, the CGL framework must be adopted when plasma is rarefied or strongly magnetized. For such plasmas, we propose the
anisotropic model (iii), for which the requirement of plasma being rarefied can be satisfied by choosing $a({\bf{r}})$
sufficiently small.

\subsection{Isotropic and anisotropic plasma equilibria in prolate spheroidal coordinates. A model of mass
exchange between two distant spheroidal objects}\label{sec_spheroids}

In this example, families of non-symmetric exact plasma equilibria is prolate spheroidal coordinates with finite magnetic
energy are obtained, in isotropic and anisotropic frameworks. On the basis of these solutions, a model of the
quasi-equilibrium stage of mass exchange by a plasma jet between two distant spheroidal objects is suggested.

\bigskip \noindent \textbf{(i). Vacuum magnetic configuration in prolate spheroidal coordinates.}
Consider the prolate spheroidal system of orthogonal coordinates:
\begin{equation}
\begin{array}{ll}
u=\theta, & 0 \leq \theta \leq \pi,\\
v=\phi, & 0 \leq \phi < 2\pi,\\
w=\eta, & 0 \leq \eta < +\infty.
\end{array}
\nonumber
\end{equation}

The coordinate surfaces are
\begin{equation}
\begin{array}{lll}
\displaystyle \frac{x^2}{a^2 \sinh^2{\eta}}+\frac{y^2}{a^2 \sinh^2{\eta}}+\frac{z^2}{a^2 \cosh^2{\eta}}=1, & ~~($prolate spheroids$, \eta=\const),\\
\displaystyle -\frac{x^2}{a^2 \sin^2{\theta}}-\frac{y^2}{a^2 \sin^2{\theta}}+\frac{z^2}{a^2 \cos^2{\theta}}=1, & ~~($two-sheet hyperboloids$, \theta=\const),\\
\displaystyle \tan(\phi)=\frac{y}{x}, & ~~($half planes$, \phi=\const),
\end{array}
\nonumber
\end{equation}

and the metric coefficients
\begin{equation}
g_{\eta\eta}=g_{\theta\theta}= a^2(\sinh^2{\eta}+\sin^2{\theta}),~~~g_{\phi\phi}=a^2 \sinh^2{\eta}\sin^2{\theta}. \nonumber
\end{equation}

It is known that the 3-dimensional Laplace equation is separable in this system, and it admits the axially-symmetric family
of solutions of this equation {\cite{moon_sp_fth}}:
\begin{equation}
\Phi=H(\eta)T(\theta),\nonumber
\end{equation}
\begin{equation}
H(\eta)=A_1 \mathcal{P}_p(\cosh{\eta}) + B_1 \mathcal{L}_p(\cosh{\eta}), ~~ T(\theta)=A_2 \mathcal{P}_p(\cos{\theta}) + B_2
\mathcal{L}_p(\cos{\theta}), \nonumber
\end{equation}
where $\mathcal{L}_p(z)\equiv\mathcal{L}^0_p(z),~\mathcal{L}_p(z)\equiv\mathcal{L}^0_p(z)$ are theLegendre wave functions of
first and second kind respectively.

\smallskip
This solution evidently satisfies both plasma equilibrium equations in magnetic-surface-related coordinates
(\ref{eq_MHD_nat_sys_1}), (\ref{eq_MHD_nat_sys_2}): the first one because the usual and truncated Laplace equations coincide
when $\Phi$ is a function of two variables, and the second - identically due to the independence of $\Phi$ on $w$.

\smallskip
In the case of integer $p$, the above 2-dimensional solution expresses in ordinary Legendre functions of the first and
second kind:
\begin{equation}
H(\eta)=A_1 P_p(\cosh{\eta}) + B_1 Q_p(\cosh{\eta}), ~~ T(\theta)=A_2 P_p(\cos{\theta}) + B_2 Q_p(\cos{\theta}).
\label{eq_prolate_massexchange}
\end{equation}

From the above family, a particular axially symmetric function $\Phi(\eta,\theta)$ with an asymptotic condition
\begin{equation}
\lim\limits_{|{\bf{r}}| \rightarrow \infty} \Phi(\eta,\theta) = M_0 z,~~~~
\end{equation}
can be chosen; its gradient is asymptotically a constant vector field in the cartesian $z$-direction:
$\grad\Phi(\eta,\theta)=M_0{\bf{e}}_z$.

This solution has the form {\cite{moon_sp_fth}}
\begin{equation}
\Phi_0(\eta,\theta) =  M_0 a \cos{\theta}\left\{n
\cosh{\eta}-\cosh{\eta_0}\frac{Q_1(\cosh{\eta})}{Q_1(\cosh{\eta_0})}\right\},\nonumber
\end{equation}
and the corresponding magnetic field ${\bf{B}}=\grad\Phi(\eta,\theta)$ is
\begin{equation}
{\bf{B}}_0 =\grad\Phi_0(\eta,\theta) =
\frac{1}{\sqrt{g_{\eta\eta}}}\frac{\partial\Phi_0(\eta,\theta)}{\partial\eta}{\bf{e}_{\eta}} +
\frac{1}{\sqrt{g_{\theta\theta}}}\frac{\partial\Phi_0(\eta,\theta)}{\partial\theta}{\bf{e}_{\theta}}.
\label{eq_B_burst_straight}
\end{equation}

\begin{figure}[tp]

\begin{center}

\vspace{0cm}

\scalebox{0.65}{\includegraphics[0cm,0cm][25cm,20cm]{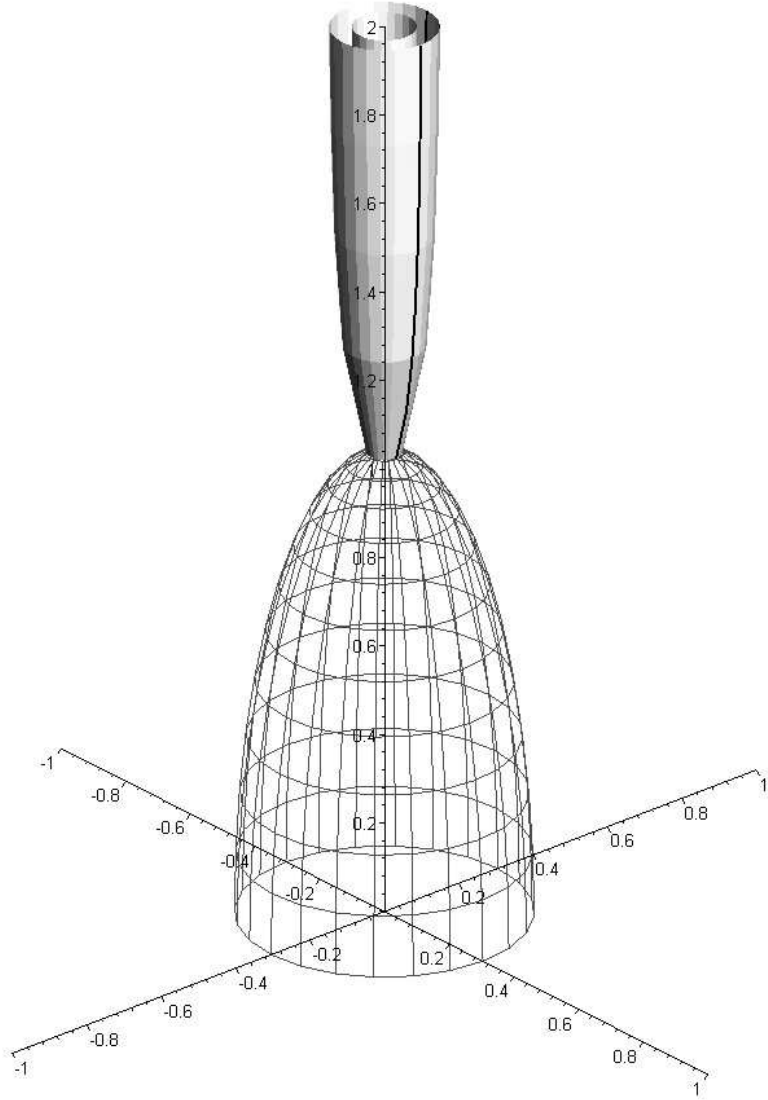}}  

\vspace{1cm}

\end{center}
\renewcommand{\baselinestretch}{1}\small\normalsize

\caption{\label{fig406}A magnetic field flux tube normal to a prolate spheroid.}{\bigskip The magnetic field
(\ref{eq_B_burst_straight}) and magnetic surfaces in prolate spheroidal coordinates.  The magnetic surfaces are nested
widening circular tubes along the $z$-axis, perpendicular to the spheroid $\eta=\eta_0$ and asymptotically approaching
circular cylinders $x^2+y^2=\const$.

The graph is built for the choice $\{a=2,~M_0=1,~\eta_0=0.3\}$. The two magnetic surfaces shown here correspond to
$\theta_1=0.07$ and $\theta_2=0.12$.

A sample magnetic field line on the outer surface ($\theta_2=0.12$) is plotted. }

\renewcommand{\baselinestretch}{1.8}\small\normalsize
\pagebreak
\end{figure}

The magnetic surfaces this field is tangent to are nested widening circular tubes along $z$-axis, perpendicular to the
spheroid $\eta=\eta_0$ and asymptotically approaching circular cylinders $x^2+y^2=\const$. A graph for the choice
$\{a=2,~M_0=1,~\eta_0=0.3\}$ with a single magnetic field line shown is presented on Fig. {\ref{fig406}}.

Each tube is uniquely defined by the value of $\theta$ of its intersection with the base spheroid. The two shown on Fig.
{\ref{fig406}} correspond to $\theta_1=0.07$ and $\theta_2=0.12$.

\bigskip
If a "winding" polar component
\begin{equation}
\frac{D}{\sqrt{g_{\phi\phi}}} {\bf{e}_{\phi}},~~~\sqrt{g_{\phi\phi}} = a \sinh{\eta}\sin{\theta},~~~D=\const, \nonumber
\end{equation}
is added to the field (\ref{eq_B_burst_straight}), which can be done by Statement {\ref{lemma_ext3}} of section
{\ref{ExtTheorems}} above, a new vacuum magnetic field is obtained:
\begin{equation}
{\bf{B}}_w =\grad\Phi_0(\eta,\theta) =
\frac{1}{\sqrt{g_{\eta\eta}}}\frac{\partial\Phi_0(\eta,\theta)}{\partial\eta}{\bf{e}_{\eta}} +
\frac{1}{\sqrt{g_{\theta\theta}}}\frac{\partial\Phi_0(\eta,\theta)}{\partial\theta}{\bf{e}_{\theta}} +
\frac{D}{\sqrt{g_{\phi\phi}}} {\bf{e}_{\phi}}. \label{eq_B_burst_winding}
\end{equation}

\begin{figure}[tp]

\begin{center}

\vspace{3cm}

\scalebox{0.6}{\includegraphics[0cm,0cm][20cm,20cm]{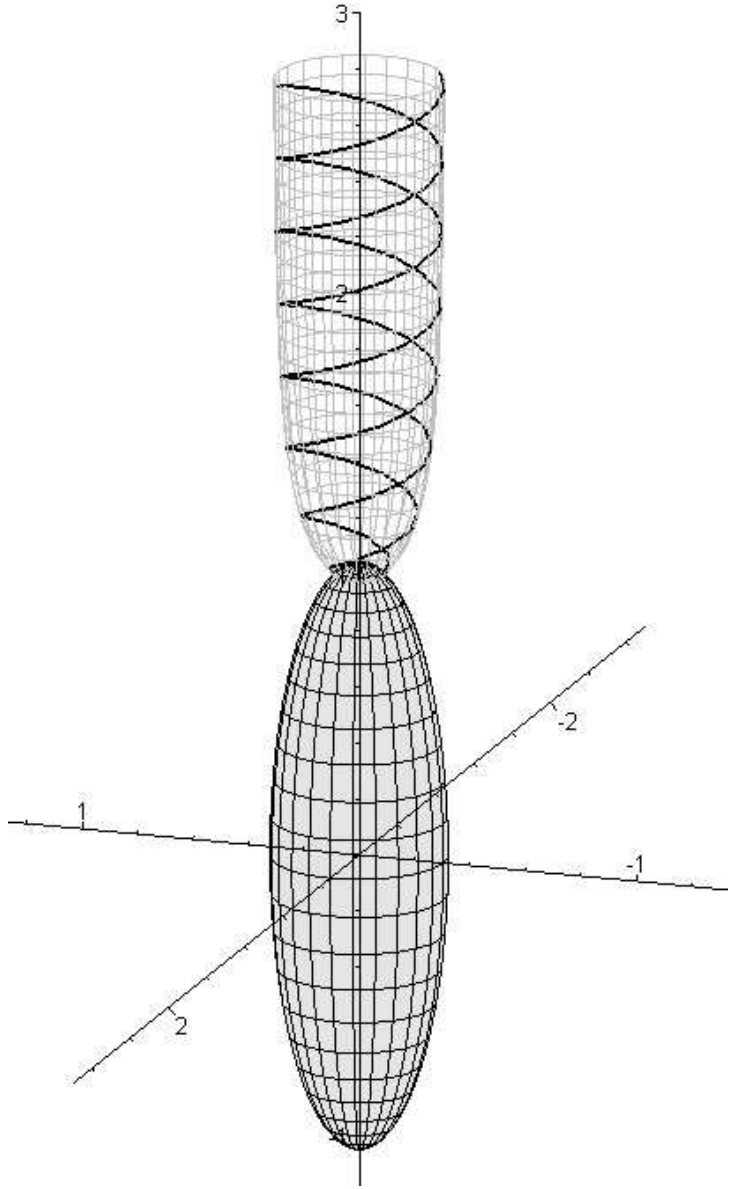}}  

\vspace{1cm}

\end{center}
\renewcommand{\baselinestretch}{1}\small\normalsize

\caption{\label{fig407}A winding magnetic field in prolate spheroidal coordinates.}{\bigskip Two opposite field lines of the
magnetic field (\ref{eq_B_burst_winding}) winding around a magnetic surface are shown; a solution is constructed in prolate
spheroidal coordinates.

The magnetic surfaces are nested widening circular tubes along the $z$-axis, perpendicular to the spheroid $\eta=\eta_0$ and
asymptotically approaching circular cylinders $x^2+y^2=\const$. The magnetic surface shown here is defined by its
intersection with the spheroid at $\theta_0=0.3$.

The graph is built for the choice $\{a=2,~M_0=1,~\eta_0=0.3\}$, $D=2.13$.}

\renewcommand{\baselinestretch}{1.8}\small\normalsize
\pagebreak
\end{figure}

A graph for $D=2.13$ showing two opposite field lines winding around a magnetic surface is presented on Fig. {\ref{fig407}}.

\bigskip \noindent \textbf{The physical model.}
We use the vacuum solutions with and without the polar component, (\ref{eq_B_burst_straight}) and
(\ref{eq_B_burst_winding}), to model a quasi-equilibrium process of \emph{mass exchange by a plasma jet between two distant
spheroidal objects}.

The useful property of the solutions is that their magnetic surfaces tend to cylinders by construction. Also, if ${\bf{B}}$
is a vacuum magnetic field, then $(-{\bf{B}})$ is a vacuum magnetic field, too.

\begin{figure}[tp]

\begin{center}

\vspace{2cm}

\scalebox{0.7}{\includegraphics[0cm,0cm][20cm,8cm]{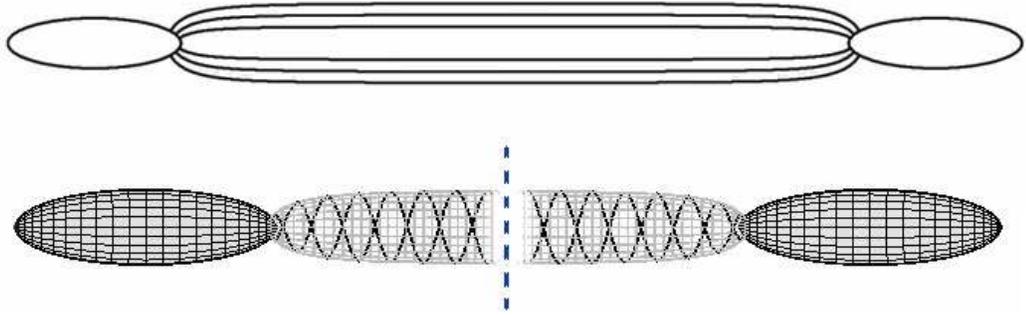}}  

\vspace{5cm}

\end{center}
\renewcommand{\baselinestretch}{1}\small\normalsize

\caption{\label{fig408}A model of mass exchange between two distant spheroidal objects by a plasma jet.}{\bigskip The
magnetic surfaces are nested widening circular tubes along the $z$-axis, perpendicular to the spheroid $\eta=\eta_0$ and
asymptotically approaching circular cylinders $x^2+y^2=\const$: $B_r/B_z = O(z^{-3})$ at $z\rightarrow \infty$.

Shown here is the procedure of gluing one copy of a solution ((\ref{eq_B_burst_winding}) or (\ref{eq_B_burst_straight}))
with another copy, rotated on the angle $\pi$ with respect to an axis orthogonal to the axis of the magnetic surface,
translated on the distance much longer than the size of the initial spheroid, and taken with the opposite sign.}

\renewcommand{\baselinestretch}{1.8}\small\normalsize
\pagebreak
\end{figure}

Hence one may effectively glue one copy of such solution with another copy, the latter being rotated on the angle $\pi$ with
respect to an axis orthogonal to the axis of the magnetic surface, translated on the distance much longer than the size of
the initial spheroid (Fig. {\ref{fig408}}), and taken with the opposite sign.

It is possible to show that for a given solution ((\ref{eq_B_burst_straight}) or (\ref{eq_B_burst_winding})), the rate of
growth of the tube radius $B_r/B_z$ has the leading term $z^{-3}$ at $z\rightarrow \infty$, hence the magnetic field lines
of the "glued" solution will not have significant "cusps" - discontinuities of derivatives.

\smallskip The resulting force-free vacuum magnetic field can be used to construct isotropic plasma equilibria with flow
by virtue of the Bogoyavlenskij symmetries (\ref{eq_OB_symm}), or anisotropic plasma equilibria with and without flow, with
the help of the MHD$\to$CGL transformations (\ref{eq_MHD_to_CGL}).

\bigskip
For example, after the application of the Bogoyavlenskij symmetries to the field ${\bf{B}}_w$, one gets an \emph{isotropic
dynamic configuration}
\begin{equation}
{\bf{B}}_1  = m({\bf{r}}) {\bf{B}}_w,~~~{\bf{V}}_1  = \frac{{n({\bf{r}})}}{{a({\bf{r}})\sqrt {\mu \rho_0({\bf{r}}) } }}
{\bf{B}}_w,\nonumber
\end{equation}
\begin{equation}
\rho _1  = a^2 ({\bf{r}})\rho_0({\bf{r}}),~~P_1  = CP_0 - n^2 ({\bf{r}}) {\bf{B}}_w^2/{(2\mu )}.
\label{eq_spheroid_model_iso}
\end{equation}
\begin{equation}
m^2 ({\bf{r}}) - n^2 ({\bf{r}}) = C = {\rm{const}},\nonumber
\end{equation}
where $a({\bf{r}}), m({\bf{r}}), n({\bf{r}}), \rho_0({\bf{r}})$ are functions constant on magnetic field lines and
streamlines (which coincide, as ${\bf{V}}_1$ and ${\bf{B}}_1$ are collinear).

To construct an anisotropic CGL extension of the above isotropic model, we again use the transformations
(\ref{eq_MHD_to_CGL}). The resulting anisotropic equilibrium ${\bf{B}}_2, {\bf{V}}_2, p_{\parallel 2}, p_{\perp 2}, \rho_2$
is then defined by ($P_0$ was set to $0$):
\begin{equation}
{\bf{B}}_2  = f({\bf{r}}) m({\bf{r}}) {\bf{B}}_w,~~{\bf{V}}_2  = g({\bf{r}}) \frac{{n({\bf{r}})}}{{a({\bf{r}})\sqrt {\mu
\rho_0({\bf{r}}) } }} {\bf{B}}_w, ~~\rho_2 = C_0 a^2 ({\bf{r}})\rho_0({\bf{r}}) \mu / g({\bf{r}})^2,\nonumber
\end{equation}
\begin{equation}
p_{\perp 2}=C_1 + \frac{{\bf{B}}_w^2}{2\mu}\left(C_0 C \mu - f^2({\bf{r}}) m^2({\bf{r}})\right),
\label{eq_spheroid_model_aniso}
\end{equation}
\begin{equation}
p_{\parallel 2}=C_1 + \frac{{\bf{B}}_w^2}{2\mu}\left( f^2({\bf{r}}) m^2({\bf{r}}) - C_0 C \mu - 2 C_0 n^2({\bf{r}})\right).
\nonumber
\end{equation}

The physical requirements and applicability bounds are the same as described in the previous model (see sec.
{\ref{sec_halfellips}}). The \emph{relation between the pressure components} of anisotropic pressure tensor, is also the
same:
\begin{equation}
\frac{p_{\perp 2}}{p_{\parallel 2}}= 1 + \frac{2 k({\bf{r}}) + 2 C_0 n^2({\bf{r}}) } {\mu f^2({\bf{r}}) m^2({\bf{r}})}
 \frac{{\bf{B}}_2^2}{2 p_{\parallel 2}},~~~k({\bf{r}})=C_0 C \mu - f^2({\bf{r}}) m^2({\bf{r}}), \nonumber
\end{equation}
which is in the agreement with the observation-based empiric formula (\ref{eq_empiric_anis1}).

\bigskip
We remark that the same way as in the previous model, the values of all the arbitrary functions of the transformations
(\ref{eq_OB_symm}), (\ref{eq_MHD_to_CGL}) can be chosen separately not on every magnetic surface, but on \emph{every
magnetic field line}. Thus these free functions are actually functions of two independent variables specifying the origin of
every magnetic line on the starting spheroid, and the resulting exact solution \emph{has no geometrical symmetries}.

\begin{figure}[tp]

\begin{center}

\vspace{2cm}

\scalebox{0.9}{\includegraphics[2cm,0cm][30cm,8cm]{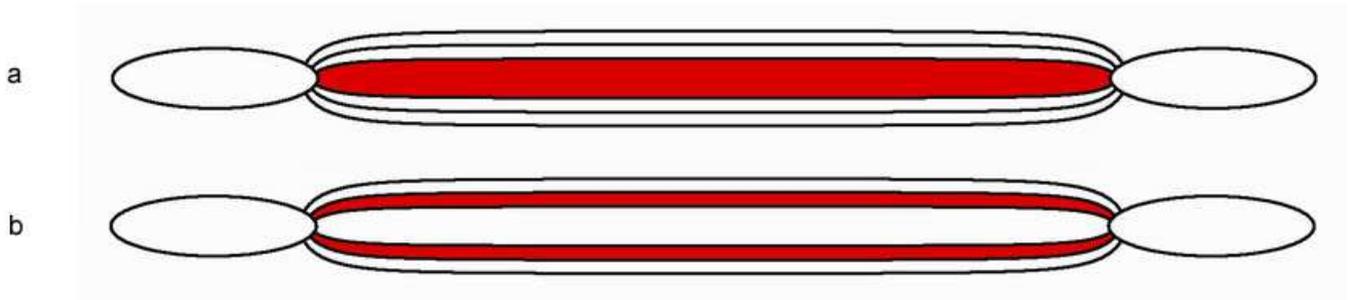}}  

\vspace{6cm}

\end{center}
\renewcommand{\baselinestretch}{1}\small\normalsize

\caption{\label{fig409}A model of mass exchange between two distant spheroidal objects by a plasma jet: possible plasma
domains.}{\bigskip Examples of possible plasma domains $\mathcal{D}$ for isotropic and anisotropic equilibria models (sec.
{\ref{sec_spheroids}}.) }

\renewcommand{\baselinestretch}{1.8}\small\normalsize
\pagebreak
\end{figure}

\bigskip
If the constant $D$ in the initial field ${\bf{B}}_w$ (\ref{eq_B_burst_winding}) is different from zero, then the $w$ -
component of this field has a singularity on the $z$-axis, and the \textbf{plasma domain} $\mathcal{D}$ must be restricted
to a volume between two nested magnetic surfaces so that the $z$-axis is excluded (see Fig. {\ref{fig409}}). However, the
families of transformed isotropic (\ref{eq_spheroid_model_iso}) and anisotropic (\ref{eq_spheroid_model_aniso}) magnetic
fields ${\bf{B}}_1$, ${\bf{B}}_2$ \emph{are smooth everywhere}, if the non-singular field ${\bf{B}}_0$
(\ref{eq_B_burst_straight}) is used instead of ${\bf{B}}_w$. Then the plasma domain $\mathcal{D}$ can be chosen to be a
region inside any flux tube or between two nested ones (Fig. {\ref{fig409}}.) The domain on Fig. {\ref{fig409}}a is simply
connected; the one on Fig. {\ref{fig409}}b is not simply connected. The axis of symmetry of the magnetic surfaces coincides
with the big axis of both spheroids.

\bigskip
In dynamic isotropic and anisotropic cases, by the properties of solutions constructed from static configurations
${\bf{V}}=0$ by the transformations (\ref{eq_OB_symm}) or  (\ref{eq_MHD_to_CGL}), the plasma velocity has the same direction
as the magnetic field, ${\bf{V}}\parallel {\bf{B}}$, so the configuration is interpreted as a magnetically driven matter
flow from one spheroid to another.

The presented solution models the quasi-equilibrium stage during the time interval $T$, with the requirement
\begin{equation}
T \cdot S_{max} \cdot \max\limits_{\mathcal{D}}|\rho{\bf{V}}| \ll M_0, \nonumber
\end{equation}
where $M_0$ is the mass of the spheroid objects, and $S_{max}$ is the area of the maximal section of the plasma domain
transverse to the flow lines.

\subsection{Example 3. Generation of orthogonal coordinate systems by coordinate transformations}\label{subs_eg_gen_cyl}

Given a coordinate system $(x^1,x^2,x^3)$ in the flat space $\mathbb{R}^3$ that satisfies Riemann equations $R_{ijkl}=0$,
one can use an arbitrary coordinate transformation
\begin{equation}
u^i=u^i(x^1,x^2,x^3),~~~i=1,2,3,\nonumber
\end{equation}
and, by tensor transformation rules, the Riemann tensor of the resulting coordinates will also be identically zero:
\begin{equation}
{R'}_{ijkl}= {R}_{abcd}\frac{\partial x^a}{\partial u^i}\frac{\partial x^b}{\partial u^j}\frac{\partial x^c}{\partial
u^k}\frac{\partial x^d}{\partial u^l}=0, ~~~i,j,k,l,a,b,c,d = 1,2,3.\nonumber
\end{equation}

Here we give an example of transformations that produce orthogonal coordinates and have metric coefficients satisfying the
sufficient condition for a force-free plasma equilibrium of the type (\ref{eq_Th_FF_gen_curv__phi1}) to exist.

\smallskip Consider the plane transformations
\begin{equation}
x=\xi_1(u,v), ~~ y=\xi_2(u,v) \label{eq_plane_transf}
\end{equation}
satisfying the Cauchy-Riemann conditions
\begin{equation}
\frac{\partial x}{\partial u}=\frac{\partial y}{\partial v}, ~~\frac{\partial x}{\partial v}=-\frac{\partial y}{\partial u}.
\label{eq_plane_CR}
\end{equation}
(Here $x,y$ are cartesian and $u,v$ curvilinear coordinates). The property of conformal mappings is that it preserves
angles, hence the families of curves $u=\const$, $v=\const$ in the plane are mutually orthogonal.

If we consider the corresponding 3D cylindrical mapping
\begin{equation}
x=\xi_1(u,v), ~~ y=\xi_2(u,v), ~~ z=w, \label{eq_cyl_transf}
\end{equation}
it defines an orthogonal coordinate system with metric coefficients
\begin{equation}
g_{11}=g_{22}=\left(\frac{\partial \xi_1(u,v)}{\partial u}\right)^2 + \left(\frac{\partial \xi_2(u,v)}{\partial u}\right)^2,
~~ g_{33}=1. \label{eq_cyl_transf_g}
\end{equation}

\bigskip \noindent \textbf{First type of solutions.}
These metric coefficients exactly satisfy the conditions used in Cases (A) and (B) of Section {\ref{SolTheorems}}, hence and
in the coordinates $(u,v,w)$ a force-free magnetic field exists:
\begin{equation}
{\bf{B}}=\left(\frac{C_1(w)}{\sqrt{g_{11}}}, \frac{C_2(w)}{{\sqrt{g_{11}}}}, 0\right), ~~\frac{\partial}{\partial
w}(C_1^2(w)+C_2^2(w))=0.\label{eq_B_FFF_cyl}
\end{equation}

Many examples of such cylindrical transformations can be suggested. The simplest ones include power, logarithmic,
exponential, hyperbolic, elliptic and other types of conformal complex plane mappings. We do not consider solutions of this
type in detail here.

\bigskip \noindent \textbf{Remark.}
The field lines of force-free magnetic fields (\ref{eq_B_FFF_cyl}) lie in planes $z=\const$. A constant $z$- component can
be added to the fields of this type as follows:
\begin{equation}
{\bf{B}}=\left(\frac{C_1(w)}{\sqrt{g_{11}}}, \frac{C_2(w)}{{\sqrt{g_{11}}}}, D\right), ~~\frac{\partial}{\partial
w}(C_1^2(w)+C_2^2(w))=0, ~~D=\const.\label{eq_B_FFF_cyl_with_z}
\end{equation}
Then the electric current density ${\bf{J}}=\curl{\bf{B}}/\mu$ does not change, and the equilibrium remains force-free.

\bigskip \noindent \textbf{Second type of solutions.}
The solutions presented below can be built in \emph{any} coordinate system obtained from the cartesian coordinates $(x,y,z)$
by a conformal plane transformation (\ref{eq_cyl_transf}).

\bigskip
In the transformed coordinates (\ref{eq_cyl_transf}), the metric coefficients are (\ref{eq_cyl_transf_g}), hence the
complete Laplace equation in coordinates $(u,v,w)$ evidently has 2-dimensional solutions $\Phi(u,v)$:
\begin{equation}
\frac{\partial^2 \Phi}{\partial u^2}+\frac{\partial^2 \Phi}{\partial v^2}=0.\nonumber
\end{equation}

Therefore, by Case (C) of Section {\ref{SolTheorems}}, in these coordinates there exists a vacuum magnetic field
(\ref{eq_B_nat_Phi}) in the $z=\const$-plane defined by
\begin{equation}
{\bf{B}}=\left(\frac{1}{\sqrt{g_{11}}} \frac{\partial \Phi}{\partial u},~~\frac{1}{\sqrt{g_{11}}} \frac{\partial
\Phi}{\partial v},~~0\right), \nonumber
\end{equation}
which can be given a non-trivial $z$-component using Statement {\ref{lemma_winding_sol}} in Section {\ref{ExtTheorems}}:
\begin{equation}
{\bf{B}}=\left(\frac{1}{\sqrt{g_{11}}} \frac{\partial \Phi}{\partial u},~~\frac{1}{\sqrt{g_{11}}} \frac{\partial
\Phi}{\partial v},~~K(u,v)\right), ~~\triangle K(u,v)=0,~~\grad\Phi(u,v)\cdot\grad K(u,v)=0.\nonumber
\end{equation}

By Lemma {\ref{lemma_winding_sol}}, such field and the pressure $P(u,v)=C-K^2(u,v)/2~~(C=\const)$ satisfy the full plasma
equilibrium system (\ref{eq_PEE})
\begin{equation}
\curl{\bf{B}}\times{\bf{B}} = \mu\grad P , ~~\div{\bf{B}}= 0.\nonumber
\end{equation}

\begin{figure}[tp]

\begin{center}

\vspace{1cm}

\scalebox{0.62}{\includegraphics[0cm,0cm][25cm,20cm]{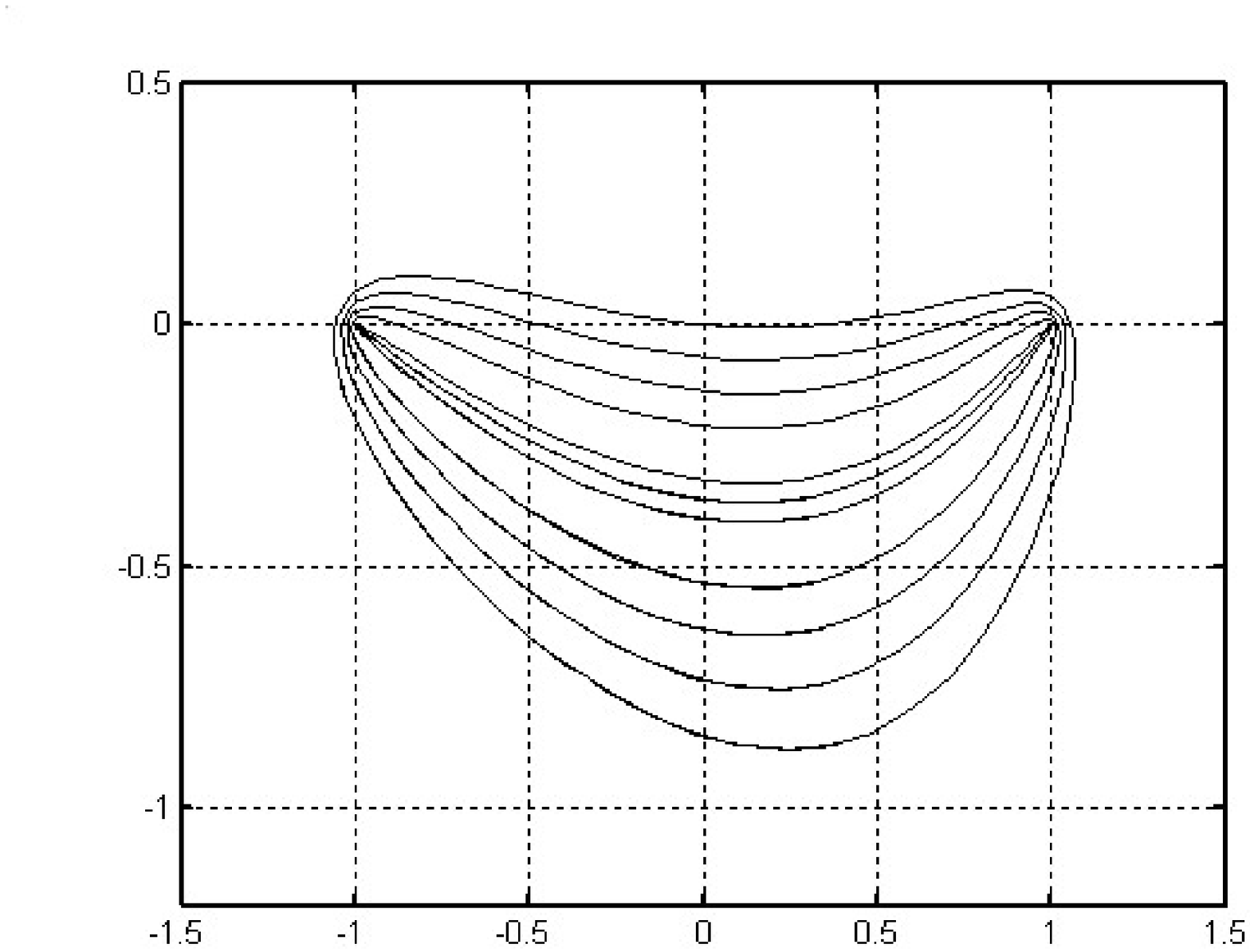}}  

\vspace{4cm}

\end{center}
\renewcommand{\baselinestretch}{1}\small\normalsize

\caption{\label{fig411}An example of non-circular cylindrical magnetic surfaces.}{\bigskip A family of cylinders with
non-symmetric closed cross-sections that are the magnetic surfaces of sample isotropic and anisotropic plasma equilibria
found in sec. {\ref{subs_eg_gen_cyl}}. }

\renewcommand{\baselinestretch}{1.8}\small\normalsize
\pagebreak
\end{figure}

\bigskip \noindent \textbf{Example.}
We now give a particular example in elliptic cylindrical coordinate system defined by a conformal transformation that acts
on the complex plane as $Z' = a \cosh Z$:
\begin{equation}
x=a\cosh{u}\cos{v}, ~~ y=a\sinh{u}\sin{v}, ~~ z=w. \nonumber
\end{equation}
We let $a=1$ and choose a function satisfying $\triangle_{(u,v)}\Phi(u,v)=0$:
\begin{equation}
\Phi(u,v)=\sinh u \cos v + 0.1 \sinh 2u \cos 2v - 3v + C_1. \nonumber
\end{equation}
A conjugate harmonic function for it is
\begin{equation}
K(u,v) =  \cosh u \sin v + 0.1\cosh 2u \sin 2v + 3 u. \nonumber
\end{equation}
The level curves $K(u,v)=\const$ are presented on Fig. {\ref{fig411}} and coincide with the projections of the magnetic
field lines on the $(x,y)$-plane.

The corresponding plasma equilibrium solution on the cylinders $K(u,v)=\const$ has a simple representation
\begin{equation}
\begin{array}{ll}
{\bf{B}}= &\displaystyle \frac{\cosh{u}\cos{v}+0.2\cosh{2u}\cos{2v}} {\sqrt{\cosh^2{u}-\cos^2{v}}} {\bf{e}_u} +
\frac{-\sinh{u}\sin{v}-0.2\sinh{2u}\sin{2v} - 3} {\sqrt{\cosh^2{u}-\cos^2{v}}} {\bf{e}_v}\\
& \displaystyle +({3u+\cosh{u}\sin{v}+0.1\cosh{2u}\sin{2v}}){\bf{e}_z}
\end{array}
\label{eq_isotr_on_cyls}
\end{equation}

\bigskip
The $u$- and $v$-components of this magnetic field evidently have a singularity at $u=v=0$ of the order $\rho^{-1}$, where
$\rho=\sqrt{u^2+v^2}$ is the "distance" to singularity. In cartesian coordinates, the singularity is located at $(x=\pm 1,
y=0)$.

The levels of the magnetic surface function $K(u,v)=\const$ encircle the singularities, and degenerate into a line segment
as $K(u,v)\rightarrow 0$, as seen from Fig. {\ref{fig411}}

\bigskip
If the magnetic field (\ref{eq_isotr_on_cyls}) with pressure $P(u,v)=C-K^2(u,v)/2$ (\ref{eq_lemma_winding_sol_pressure}) is
to be used in any \textit{isotropic} model, the plasma domain $\mathcal{D}$ is to be restricted to the cylindrical volume
between any two magnetic surfaces: $\mathcal{D}=\{(u,v): 0<K_1\leq K(u,v)\leq K_2$. Using the fact that the magnetic field
is tangent to the surfaces $K(u,v)=\const$, this can be done by introducing a boundary surface current
(\ref{eq_bdry_current}). Then outside of $\mathcal{D}$ magnetic field is zero: ${\bf{B}}\equiv 0$.

\bigskip
However, if one is to use the described configuration as an initial solution for building an \textit{anisotropic static
equilibrium} by virtue of the transformations (\ref{eq_MHD_to_CGL}), then he should select the function $f({\bf{r}})$ (that
must be constant on magnetic surfaces, $f({\bf{r}})=f(K)$) as follows:
\begin{equation}
f(K)=\frac{1}{\max\limits_{K(u,v)=K}\{|B_1|,|B_2|\}} a(K),\nonumber
  \end{equation}
where $a(K)$ is some function with compact support and the property ${\frac{da(K)}{dK}}|_{K=0} =0$, and $|B_1|,|B_2|$ are
respectively the $u$- and $v$-components of the field (\ref{eq_isotr_on_cyls}).

This function is evidently finite in any domain $\mathcal{D}$ bounded by a level $K(u,v)\leq K$. On the degenerate surface
$K(u,v)=0$ by continuity we have ${\bf{B}}=0$. On the outer boundary of the domain, a surface current
(\ref{eq_bdry_current}) must be introduced, to ensure so that ${\bf{B}}\equiv 0$ outside of $\mathcal{D}$.

\bigskip \noindent \textbf{Remark. Vacuum magnetic fields in rotational coordinate systems.}

Among the classical and esoteric  coordinate systems where the Laplace equation is separable or R-separable
{\cite{moon_sp_fth}}, many are rotationally symmetric systems, with metric coefficients independent of the polar angle
$\phi$.

In all such systems, the Laplace equation has solutions independent of $\phi$. Examples are toroidal coordinates, usual and
inverse prolate and oblate spheroidal coordinates, cap-cyclide, disk-cyclide, cardioid coordinates and several others (see
{\cite{moon_sp_fth}}).

By the statement of Case (C) of Section {\ref{SolTheorems}}, gradient "vacuum" magnetic fields can be built in such
coordinates, tangent to the magnetic surfaces, which are in this case vertical half-planes $\phi=\const$. Again, a non-zero
$\phi$-component can be added to these vacuum magnetic fields, to make them non-planar.

Examples of such vacuum magnetic fields and the corresponding isotropic and anisotropic plasma configurations obtained from
them by transformations (\ref{eq_OB_symm}), (\ref{eq_MHD_to_CGL}) will be built in consequent papers.

Particular solutions obtained in different coordinates can have simple algebraic representation only in the corresponding
coordinates, therefore different rotational coordinate systems may not be considered equivalent, from the computational
point of view.

\smallskip We also remark that the magnetic fields constructed this way \emph{can not} be found from Grad-Shafranov equation,
which describes plasma equilibria with magnetic surfaces $\Psi(r,z)=\const$, whereas in the above case magnetic surfaces are
$\Psi=\Psi(\phi)=\const.$

\section{Conclusion}\label{sec_concl}

In this paper, a method of construction of exact plasma equilibria is presented. It is used for producing dynamic and static
equilibria in different geometries, in both classical MHD and anisotropic tensor-pressure CGL frameworks.

The method is based on representing the system of static isotropic plasma equilibrium equations (\ref{eq_PEE}) in
coordinates $(u,v,w)$, such that magnetic surfaces coincide with the coordinate level surfaces $w=\const$. Such
representation is valid when the family of magnetic surfaces can be a part of a triply orthogonal system of surfaces, i.e.
when it forms \emph{a family of Lam\'{e}}. In such coordinates, the system of four static Plasma Equilibrium equations
(\ref{eq_PEE}) is reduced to two partial differential equations for two unknown functions. The first of the equations of the
system is a "truncated" Laplace equation (\ref{eq_MHD_nat_sys_1}), and the second one, (\ref{eq_MHD_nat_sys_2}), has an
energy-connected interpretation (Section {\ref{GeomSec}}.)

Instead of four unknown functions of the static MHD equilibrium system, ${\bf{B}}({\bf{r}})$ and $P({\bf{r}})$, that depend
on three spatial variables, the new system of equations employs only two functions - $\Phi(u,v,w)$ and $P(w)$, and the
magnetic field ${\bf{B}}({\bf{r}})$ is reconstructed from the relation (\ref{eq_B_nat_Phi}).

\bigskip
In Section {\ref{AppPropSec}}, sufficient conditions on the metric coefficients are established under which exact solutions
of particular types can be found in corresponding coordinates. In particular, if the conditions (\ref{eq_Th_FF_gen_curv1})
or (\ref{eq_Th_FF_gen_curv2}) are satisfied, a force-free plasma equilibrium ($P(w)=\const$) can be constructed.

In coordinates where the general 3D Laplace equation admits 2-dimensional solutions, "vacuum" magnetic fields
$\div{\bf{B}}=0$,~$\curl{\bf{B}}=0$  tangent to magnetic surfaces $w=\const$) can be built; such field often have
non-trivial geometry. Statements {\ref{lemma_winding_sol}} and {\ref{lemma_ext3}} extend this class of solutions in some
coordinate systems.

\smallskip
The use of "vacuum" gradient fields for plasma equilibrium modeling is discussed. Such solutions can serve as initial
solutions in the infinite-parameter Bogoyavlenskij symmetries (\ref{eq_OB_symm}), or in the MHD $\to$ CGL transformations
(\ref{eq_MHD_to_CGL}). The application of these symmetries and transformations, for each initial solution, generates
families of dynamic MHD and static and dynamic CGL equilibria; suitable solutions from these families can be used as
physical models.

\bigskip
We use the described procedure to construct particular examples of plasma equilibria. The first example (Section
{\ref{SolTheorems}}) is a set of non-Beltrami Force-Free plasma equilibria (\ref{eq_FF}) in a prescribed geometry - with
spherical magnetic surfaces and the force-free coefficient $\alpha({\bf{r}})$ (\ref{eq_FF}) being a function of the
spherical radius. When used in a linear combination, these solutions give rise to force-free fields with \emph{no
geometrical symmetries} tangent to spheres.

\smallskip
In the second example (Section {\ref{sec_halfellips}}), we build exact dynamic isotropic and anisotropic plasma equilibrium
configurations with magnetic fields tangent to \emph{ellipsoids}. We start from finding a set of non-trivial "vacuum"
magnetic fields tangent to ellipsoids, which are then transformed into families of dynamic isotropic and anisotropic plasma
equilibria by virtue of Bogoyavlenskij symmetries (\ref{eq_OB_symm}) and "anisotropizing" transformations
(\ref{eq_MHD_to_CGL}). These solutions are well-defined and have a finite magnetic energy in half-space. They model solar
coronal flares near the active regions of the Sun photosphere. The resulting anisotropic model is an essentially
non-symmetric, unlike other available models (see {\cite{bisk}}.) It reproduces the features of solar flares known from
observations, including the presence of thin current sheets.

\smallskip
In the third example (Section {\ref{sec_spheroids}}), the coordinate representation is used to build a particular trivial
"vacuum" magnetic field in prolate spheroidal coordinates. From that field, also by the symmetries (\ref{eq_OB_symm}) and
the transformations (\ref{eq_MHD_to_CGL}), we construct families of non-degenerate (and generally non-symmetric) isotropic
and anisotropic plasma equilibria with dynamics, which model the quasi-stationary phase of mass exchange between two
spheroidal objects. Plasma domains of different geometry and topology may be chosen for the model.

\smallskip
In the subsection {\ref{subs_eg_gen_cyl}}, conformal coordinate transformations $(x,y,z)\rightarrow(u,v,w)$ are employed to
generate orthogonal coordinates $(u,v,w)$, where exact plasma equilibria can be constructed using formulas from Section
{\ref{SolTheorems}}. A family of plasma equilibrium configurations with non-circular cylindrical magnetic surfaces and
realistic values of plasma parameters is obtained.

\bigskip
The suggested approach can be used for the construction of fully 3-dimensional static and dynamic plasma equilibrium
solutions for MHD and CGL continuum plasma models. Configurations with and without geometrical symmetries, for plasma
domains of different shapes, can be found.

\bigskip
The author thanks Dr. Kayll Lake and Dr. Oleg Bogoyavlenskij for discussion, and the Natural Sciences and Engineering
Research Council of Canada (NSERC) for research support.


\end{document}